\documentclass[11pt,letterpaper]{article}
\usepackage[T1]{fontenc}
\usepackage[utf8]{inputenc}
\usepackage{lmodern}
\usepackage{textcomp}
\usepackage{amsmath,amssymb}
\usepackage[margin=1in]{geometry}
\usepackage{graphicx}
\usepackage{booktabs}
\usepackage{array}
\usepackage{longtable}
\usepackage{tabularx}
\usepackage{float}
\usepackage{seqsplit}
\usepackage{needspace}
\usepackage{ragged2e}
\usepackage[font=small,labelfont=bf,labelformat=empty,skip=5pt,width=0.94\linewidth]{caption}
\usepackage{mdframed}
\usepackage{microtype}
\usepackage[section]{placeins}
\usepackage[hidelinks,breaklinks=true]{hyperref}
\usepackage{xurl}
\hypersetup{pdftitle={Clean Scores, Buried Evidence, and Confident Wrong: A Receipt-Based Audit of Frontier Agentic QA},
            pdfauthor={Luis M. Sanchez},pdfsubject={Agentic QA evaluation; receipts; calibration},
            pdfkeywords={LLM agents, benchmark, retrieval, calibration, audit}}

\newcolumntype{R}{>{\raggedleft\arraybackslash}X}
\newcolumntype{L}{>{\raggedright\arraybackslash}X}

\newcommand{\paperSection}[1]{%
  \phantomsection\addcontentsline{toc}{section}{#1}%
  \section*{#1}}
\newcommand{\paperSubsection}[1]{%
  \phantomsection\addcontentsline{toc}{subsection}{#1}%
  \subsection*{#1}}

\newcommand{\appendixDivider}[1]{%
  \phantomsection\addcontentsline{toc}{section}{#1}%
  \begin{center}{\LARGE\bfseries #1}\\[6pt]\rule{0.3\linewidth}{0.6pt}\end{center}%
  \vspace{6pt}}
\newcommand{\appendixSection}[2]{%
  \FloatBarrier\Needspace*{0.3\textheight}\phantomsection\addcontentsline{toc}{section}{#1 --- #2}%
  \vspace{14pt plus 4pt minus 2pt}%
  {\noindent\rule{\linewidth}{0.5pt}\par\vspace{4pt}%
   \noindent{\large\bfseries #1}\hspace{0.9em}{\large\bfseries #2}\par}\vspace{8pt}}
\newcommand{\appendixSectionFirst}[2]{%
  \phantomsection\addcontentsline{toc}{section}{#1 --- #2}%
  \vspace{14pt plus 4pt minus 2pt}%
  {\noindent\rule{\linewidth}{0.5pt}\par\vspace{4pt}%
   \noindent{\large\bfseries #1}\hspace{0.9em}{\large\bfseries #2}\par}\vspace{8pt}}
\newcommand{\boxheading}[1]{%
  \phantomsection\addcontentsline{toc}{section}{#1}%
  {\noindent\large\bfseries #1\par}\vspace{2pt}}

\mdfdefinestyle{boxone}{linewidth=0.8pt,linecolor=black,%
  innertopmargin=8pt,innerbottommargin=8pt,innerleftmargin=10pt,%
  innerrightmargin=10pt,skipabove=8pt,skipbelow=8pt}

\title{\bfseries Clean Scores, Buried Evidence, and Confident Wrong: A Receipt-Based Audit of Frontier Agentic QA}
\author{Luis M. S\'anchez\\ Toryx Inc.\\ \texttt{luis.m.sanchez@toryx.ai}}
\date{}

\begin{document}
\maketitle

\begin{abstract}
\noindent Frontier models score well on shallow document/chart reading tasks. In a controlled data-room audit, moving evidence into buried conditions reduced accuracy, increased forced declarations, increased tool calls, and increased cost per correct answer. Confidence and benchmark calibration did not fully capture wrong answers; a documented production incident shows fabricated structural claims can be mixed with accurate numeric tables. Agentic evaluations need claim-level receipts (statement-level provenance, not answer-level scores), condition-aware scoring, and human-adversarial verification --- an auditing discipline, not a leaderboard. The accuracies reported here are produced by a frozen containment grader rather than an exact-match one; a structured audit of that grader, including blinded human review, found several conclusions sensitive to it, and we state throughout which survive that sensitivity and which do not. The setting we measure is financial due diligence; the setting we are building toward next is defense staff work, where the same buried-evidence shape appears. In both, the model is not a party to the consequences; the person who signs is. In plain terms: in the documented cases we examine, agents can pair accurate numbers with confident fabricated explanations, and the burden of proof must therefore move from the model to the evidence trail. These findings motivate evaluating the complete evidence-to-answer pipeline rather than the reader alone. A central open question is whether explicit, source-linked evidence structures and training on audited task trajectories can improve evidence delivery and answer accuracy under matched resource budgets, including for open-weight models. The present study establishes the baseline and auditing requirements for that investigation; it does not evaluate those interventions.
\end{abstract}

\paperSection{1. Introduction}

Large language models are increasingly deployed as \textit{agents}: given a folder of documents and a set of tools, they are asked to find, combine, and report facts the way a junior analyst would. Benchmarks, however, usually hand the model a clean, curated context --- the equivalent of handing the analyst only the files that contain the answer. Real work is not like that. Evidence is buried among noise, and the cost of finding it is real.

The shape of that work is the same across regulated settings. A credit associate asked whether a fund may add to a position must find the mandate version in force on the meeting date, the concentration clause inside it, and the current NAV in a month-end report --- three documents, none of which name each other. A contracting officer asked whether an offeror's supply chain is clean must extract the Tier-1 to Tier-3 suppliers from the proposal, check each against Section 889, UFLPA and OFAC lists, trace ownership to beneficial owners, and cite the controlling entry for each flag. In both cases the answer is only as good as the chain behind it, and in both cases the person who signs it is accountable for that chain.

That asymmetry is the reason for this paper. A judge advocate who clears a target list, an analyst whose assessment feeds a national-security decision, a portfolio manager who buys a security for a pension fund --- each signs, and each answers for what they signed. If the chain behind the answer was assembled by a model that is right between roughly one-third and two-thirds of the time at depth five (Table 4: buried accuracy at d5 runs from 38\% to 71\% across the panel, and we have not measured beyond that), and that states the same confidence whether it is right or wrong (\S{}6), then the signature is a wager. \textit{The model is not a party to the wager. It faces no court-martial, no enforcement action, no fiduciary claim; the human does. That is why this paper insists on receipts rather than confidence: a receipt is the one thing a signer can check before signing, and the one thing that turns ``do I trust the model?'' into ``can I verify the chain?''}

This paper measures the first setting; \S{}9 describes how the same instrument is being carried to the second.

This paper asks a simple question: \textbf{what happens to frontier agents when the evidence is buried?} We build a synthetic finance-and-insurance ``data room'' --- a firewalled folder of documents (filing-style prints, credit memos, committee minutes, mandates, market-event notices) that we authored so every fact has known provenance --- and run one flagship model from each of six leading labs (OpenAI, Anthropic, Zhipu, Alibaba, Google, Meta) under two conditions: \textit{clean} (only the answer-bearing documents mounted) and \textit{buried} (the full room, including noise). We measure accuracy, committed-error rate, forced declarations, tool use, and cost --- a multi-metric, cost-aware view in the spirit of holistic evaluation [1] and of recent cost-per-benchmark accounting [2]. Existing agent benchmarks score task completion on open-web or software tasks [3, 4, 5]; multi-hop QA datasets test composition over open corpora [6, 7]; and long-context studies show that retrieval degrades with where the evidence sits [8, 9]. None combines a closed, provenance-known corpus with a claim-level audit, which is the gap this instrument fills.

Because a model can produce a confident, well-formatted answer that is wrong --- or even pair accurate numbers with fabricated explanations [10, 11] --- we treat every claim as needing a \textit{receipt}: a trace to the specific evidence that supports it. A receipt is the minimal, verifiable form of an explanation --- it does not say \textit{why} the model answered, only \textit{on what}, and that is the property an auditor can check; it is the auditability requirement of the trustworthy-AI literature [12, 13] reduced to its checkable core. We report what the panel does and does not do, and we are explicit about what we do not claim.

A wrong answer under buried evidence does not, by itself, identify the failing component. The agent may not retrieve sufficient evidence, may retrieve it without receiving the relevant passage in its usable context, may misinterpret or combine the evidence incorrectly, or may fail the answer-format contract. The evaluator may also reject a substantively valid answer. Distinguishing these cases matters because they imply different remedies: retrieval improvements, changes to evidence presentation, reader training, interface corrections, or repairs to the evaluation instrument. This paper motivates that decomposition and supplies the instrument for it; it does not establish which component dominates.

\paperSection{2. Scope and non-claims}

This report presents a controlled audit of agentic question answering, not a general leaderboard. We selected one flagship model from each of six leading laboratories --- OpenAI, Anthropic, Zhipu, Alibaba, Google, and Meta --- the major closed frontier labs (the leading open-weight release is excluded as a different category; Grok-4.6 appears in the shallow quiz, Table 1b, and the Wave-1 appendix, Table A.1; Nemotron-3-Ultra-550B appears only as an excluded run in Table A.1; both are outside the panel). From this point, tables and figures name both the laboratory and the model as run (Lab · Model); prose refers to laboratories, with the model behind each given in Table 1. The panel is a design choice (one flagship per lab), not a claim about which models are best.

\begin{table}[htbp]
\centering
\caption*{\textbf{Table 1.} The six-lab panel: one flagship model per lab, with this paper's own measured room02 accuracy (from Table 2).}
\small
\begin{tabular}{llrr}
\toprule
\textbf{Lab} & \textbf{Flagship model (as run)} & \textbf{room02 clean acc} & \textbf{room02 buried acc} \\
\midrule
OpenAI & GPT-5.6-sol & 82.9\% & 78.9\% \\
Anthropic & Fable 5 & 89.4\% & 77.9\% \\
Zhipu & GLM-5.3 & 89.4\% & 80.5\% \\
Alibaba & Qwen3.8-Max\textsuperscript{\ref{fn:qwen}} & 87.0\% & 74.0\% \\
Google & Gemini-3.1-Pro & 85.2\% & 67.5\% \\
Meta & Muse-Spark 1.2 & 82.9\% & 74.0\% \\
\bottomrule
\end{tabular}
\end{table}

\FloatBarrier

\textit{Table 1 notes.} Reported or rumored parameter counts, context windows and the effective \$/M-token rates of this run (clean-arm spend $\div$ clean-arm tokens, not list price) are recorded in the frozen pricing snapshot (\path{configs/openrouter_pricing_2026-09-04.json}) and are deliberately not reproduced here, because they are not independently verified. The accuracy columns are this paper's own measured benchmark (Table 2), which is verified and frozen --- we do not publish external benchmark scores we have not reproduced. \footnote{\label{fn:qwen}Qwen3.8-Max is priced at OpenRouter's published per-token rate; in practice the author accessed the model through Alibaba's cloud coding-plan (flat-rate) subscription. The same convention applies wherever a Qwen3.8-Max cost appears (Tables 3 and 3b).}

Throughout this paper, multi-hop labels are reported as ``declared hop count'' rather than proven compositional depth. The prompt audit (Appendix A) found that 0 of 41 room02 prompts enumerate substeps or hand the model its decomposition; hop labels therefore remain corpus-declared task complexity, with the caveat that depth is inferred from corpus design. A documented fabrication case involving Fable 5 (Box 1) is included strictly as an n=1 production observation; it is not used to estimate general model-level fabrication rates.

\paperSection{3. Instruments and receipts}

A ``data room'' is a firewalled folder of documents mounted read-only for the model, which acts as an analyst agent --- the work of a junior analyst or senior associate at a hedge fund or boutique investment bank, spanning credit, equities, and commodities --- with shell tools (bash, python, read, grep) --- retrieval is performed by the agent's own tool calls, not by a fixed retriever as in retrieval-augmented generation [14]. room02 is our second-generation room: a synthetic finance-and-insurance document corpus --- public-filing-style quarterly and annual prints, credit memos, investment-committee minutes, mandate documents, and market-event notices (sanctions, ratings, index changes, tender offers) --- authored for this project so that every fact has known provenance and no real-world proprietary material is used in the corpus (unlike financial QA sets built from public filings [15]); model tool-output logs (\texttt{stdout\_head}) can echo public SEC filing text, which is why the raw logs are excluded from the planned public deposit (Appendix D.4).

\paperSubsection{room02 data-room benchmark}

The primary instrument is the room02 data-room benchmark, consisting of 41 synthetic items. The items were designed and curated by human experts to reflect real-life scenarios encountered in these financial settings. The items have a declared hop count ranging from 2 to 5, distributed as follows: d2=8, d3=9, d4=17, d5=7. The benchmark was run across 3 seeds per model-condition.

The evaluation harness mounts a read-only data room at \texttt{/work/documents} and provides the model with tool access, specifically \texttt{bash}, \texttt{python}, \texttt{read\_file}, \texttt{grep}, and \texttt{find}. Two conditions are evaluated: a ``clean'' condition where only gold documents are mounted, and a ``buried'' condition where the full room, including noise, is mounted. Clean and buried differ in the mounted room contents; per-condition input-token ceilings also differ but were non-binding (the largest observed input was below both ceilings), so no arm was truncated by its ceiling.

Answers were elicited with a forced-declaration JSON schema requiring an answer, a confidence score from 0 to 100, and a list of evidence files. Models are evaluated on accuracy, confident-wrong rate, forced declarations, tool calls, API spend, and cost per correct answer.

\paperSubsection{Shallow chart-reading instrument (image input)}

\textit{Multimodal instrument.} Regulated work is not text-only. An insurance adjuster's agent reads dash-cam and collision video; a targeting cell reads satellite and infrared imagery; a credit analyst reads a chart pasted into a board deck or listens to a roadshow recording that never gets transcribed. The room02 data room deliberately isolates the text case, but the paper carries one multimodal instrument alongside it: a shallow chart-and-table quiz in which every item is an image (a chart rendered as a terminal screenshot or a newspaper scan) and the model must read a value off the picture before reasoning with it. It is intentionally 1- and 2-hop, so that it measures reading, not composition. Two things it shows already: most image-capable models score 100\% at one hop and slightly lower at two; Google is the exception at 67.6\% and 92.6\% (Table 1b). Not every panel member took this instrument --- GLM-5.3 was run in the text-based document experiment but not here, because the configuration we tested did not accept image input --- so the quiz covers five of the six panel labs, and the two instruments should not be read as a matched pair. The multimodal case is where the author expects the buried-evidence effect to be largest, because the evidence is not merely buried in a folder but encoded in a modality the model reads worst; it is the subject of forthcoming work (\S{}9).

To establish a commodity baseline on this instrument, models were evaluated on the quiz as follows. The frozen wave-1 file \texttt{chart\_v2\_main.jsonl} (612 rows) holds six model strings, of which only three belong to the room02 panel (OpenAI, Anthropic, Google); Alibaba's and Meta's flagships were added in a wave-2 addendum run on 2026-09-10 with the same items, variants and seed (\path{chart_v2_main_wave2.jsonl}, 204 scored rows, frozen with its own manifest; Appendix D.3), and Zhipu's flagship was not run on this instrument because the configuration we tested did not accept image input. Table 1b reports the flagship, closed-weight subset under the rule in \path{make_table2_quiz_baseline.py} (the wave-1 generator, whose rule the wave-2 script \path{make_table2_quiz_baseline_wave2.py} inherits; Appendix D.2): it drops Gemini 2.5 Pro (not the Google flagship) and Qwen3.8-27B (open-weight, internal scouting only); their rows are retained in the frozen file but are not reported. Grok-4.6 appears although it is outside the panel. This instrument is deliberately shallow (1- and 2-hop); deeper chains (3--5 hops) are measured in the room02 data room (Table 4), not here. Five of the six image-tested labs score 100\% at one hop and lose 4--9 points at two hops (91.2--95.6\%); of the five panel labs that could take the quiz, four score 100\% and Google is the exception, at 67.6\% on one hop and 92.6\% on two. The instrument comprises 34 underlying questions, each run in three variants, yielding 34 one-hop and 68 two-hop scored rows: 102 rows per model at seed 0.

\begin{table}[htbp]
\centering
\caption*{\textbf{Table 1b.} Shallow chart-reading quiz (image input): accuracy by hop and combined. 34 underlying questions, each run in three variants: 34 one-hop scored rows and 68 two-hop scored rows, 102 rows per model at seed 0. Five of the six room02 panel labs, plus Grok-4.6 as a supplementary model outside the panel. GLM-5.3 was evaluated in the text-based document experiment but not in this image-input quiz because the tested configuration did not support images. Rows for Fable 5, GPT-5.6-sol, Grok-4.6 and Gemini-3.1-Pro are the frozen wave-1 run; Qwen3.8-Max and Muse-Spark 1.2 are the wave-2 addendum (Appendix D.3).}
\small
\begin{tabular}{llrrrr}
\toprule
\textbf{Lab} & \textbf{Model} & \textbf{1-hop acc} & \textbf{2-hop acc} & \textbf{Overall acc} & \textbf{Mean stated conf.} \\
\midrule
Anthropic & Fable 5 & 100.0\% & 95.6\% & 97.1\% & 89.3 \\
OpenAI & GPT-5.6-sol & 100.0\% & 95.6\% & 97.1\% & 97.9 \\
Alibaba & Qwen3.8-Max & 100.0\% & 94.1\% & 96.1\% & 89.9 \\
Meta & Muse-Spark 1.2 & 100.0\% & 92.6\% & 95.1\% & 93.6 \\
xAI & Grok-4.6 & 100.0\% & 91.2\% & 94.1\% & 88.7 \\
Google & Gemini-3.1-Pro & 67.6\% & 92.6\% & 84.3\% & 97.8 \\
\bottomrule
\end{tabular}
\end{table}

\paperSubsection{Receipts}

This audit relies on receipts to verify model behavior. The buried data-room condition tests retrieval under friction. Claim-level receipts trace each claim to a source file. The intent is traceability in the trustworthy-AI sense: a third party with the frozen pack and no access to the model should be able to reconstruct which document supported which statement.

\paperSection{4. Main results: room02 clean vs buried}

In the controlled data-room agentic benchmark, moving evidence from clean to buried conditions changes accuracy, forcing behavior, tool-call volume, and cost.

\textit{How to read the accuracy columns.} Every accuracy figure in this and the following tables is a \textbf{frozen historical score over all 41 items}, produced by the containment-based grader in force when the results were frozen. These are \textbf{not exact-match scores}, and \S{}8 describes what an audit of that grader found. The statistical comparisons in \S{}8 use a restricted \textbf{39-item} population, two items having been ruled defective against their own source documents; that population is used only there and the tables below are not restated.

\begin{table}[htbp]
\centering
\caption*{\textbf{Table 2.} room02 accuracy \& confidence (clean vs buried).}
\setlength{\tabcolsep}{3pt}
\footnotesize
\begin{tabular}{lllrrrrr}
\toprule
\textbf{Lab} & \textbf{Model} & \textbf{Condition} & \textbf{Acc \%} & \textbf{CW \%} & \textbf{Forced \%} & \textbf{Acc ans. (0--1)} & \textbf{Acc forc. (0--1)} \\
\midrule
OpenAI & GPT-5.6-sol & clean & 82.9\% & 17.1\% & 0\% & 0.829 & --- \\
OpenAI & GPT-5.6-sol & buried & 78.9\% & 9.8\% & 20\% & 0.878 & 0.440 \\
Anthropic & Fable 5 & clean & 89.4\% & 10.6\% & 0\% & 0.894 & --- \\
Anthropic & Fable 5 & buried & 77.9\% & 5.7\% & 55\% & 0.870 & 0.716 \\
Zhipu & GLM-5.3 & clean & 89.4\% & 9.8\% & 2\% & 0.900 & 0.667 \\
Zhipu & GLM-5.3 & buried & 80.5\% & 19.5\% & 0\% & 0.805 & --- \\
Alibaba & Qwen3.8-Max & clean & 87.0\% & 13.0\% & 0\% & 0.870 & --- \\
Alibaba & Qwen3.8-Max & buried & 74.0\% & 26.0\% & 0\% & 0.740 & --- \\
Google & Gemini-3.1-Pro & clean & 85.2\% & 14.8\% & 0\% & 0.852 & --- \\
Google & Gemini-3.1-Pro & buried & 67.5\% & 32.5\% & 0\% & 0.675 & --- \\
Meta & Muse-Spark 1.2 & clean & 82.9\% & 17.1\% & 0\% & 0.829 & --- \\
Meta & Muse-Spark 1.2 & buried & 74.0\% & 25.2\% & 3\% & 0.740 & 0.750 \\
\bottomrule
\end{tabular}
\end{table}

\FloatBarrier

\textit{Table 2 notes.} Acc = share of scored rows correct. Each arm planned 123 item-seeds (41 items $\times$ 3 seeds); scored n is 122 for Anthropic buried, 122 for Google clean, 117 for Meta clean, and 123 elsewhere (Table 4 notes give the exclusions). Anthropic buried includes one incorrect \texttt{budget\_exhausted} row: it counts in Acc but in neither conditional accuracy nor CW. CW\% = committed error: rows answered freely that were wrong, over all scored rows. Forced = share of rows answered under a closed search budget (graded normally, so Acc forced is nonzero). Acc answered / Acc forced = accuracy restricted to each row class, reported as fractions (0--1); Acc, CW and Forced are percentages.

\begin{table}[htbp]
\centering
\caption*{\textbf{Table 3.} room02 cost \& effort (clean vs buried).}
\small
\begin{tabular}{lllrrrr}
\toprule
\textbf{Lab} & \textbf{Model} & \textbf{Condition} & \textbf{\$} & \textbf{\$/correct} & \textbf{\$/wrong} & \textbf{Mean tool calls} \\
\midrule
OpenAI & GPT-5.6-sol & clean & \$5.63 & \$0.055 & \$0.268 & 5.8 \\
OpenAI & GPT-5.6-sol & buried & \$36.09 & \$0.372 & \$1.388 & 18.4 \\
Anthropic & Fable 5 & clean & \$12.94 & \$0.118 & \$0.995 & 7.0 \\
Anthropic & Fable 5 & buried & \$77.97 & \$0.821 & \$2.888 & 16.1 \\
Zhipu & GLM-5.3 & clean & \$11.92 & \$0.108 & \$0.917 & 9.2 \\
Zhipu & GLM-5.3 & buried & \$58.43 & \$0.590 & \$2.434 & 25.7 \\
Alibaba & Qwen3.8-Max\textsuperscript{\ref{fn:qwen}} & clean & \$17.71 & \$0.166 & \$1.107 & 11.7 \\
Alibaba & Qwen3.8-Max\textsuperscript{\ref{fn:qwen}} & buried & \$70.74 & \$0.777 & \$2.211 & 25.3 \\
Google & Gemini-3.1-Pro & clean & \$18.81 & \$0.181 & \$1.045 & 9.8 \\
Google & Gemini-3.1-Pro & buried & \$99.31 & \$1.197 & \$2.483 & 22.9 \\
Meta & Muse-Spark 1.2 & clean & \$38.31 & \$0.395 & \$1.915 & 12.6 \\
Meta & Muse-Spark 1.2 & buried & \$55.71 & \$0.612 & \$1.741 & 19.9 \\
\bottomrule
\end{tabular}
\end{table}

\FloatBarrier

\textit{Table 3 notes.} \$ = total spend for the arm; \$/correct and \$/wrong = spend divided by correct / wrong answers (being wrong costs real retrieval effort). Mean tool calls = average tool invocations per item. \$/M-token prices are prevailing market rates as of August 2026. See footnote~\ref{fn:qwen}.

\begingroup\small
\begin{center}\begin{minipage}{\linewidth}\captionsetup{type=table}\caption*{\textbf{Table 3b.} Clean $\rightarrow$ buried deltas per lab (from Tables 2 and 3).}\end{minipage}\end{center}
\noindent\begin{tabularx}{\linewidth}{llRRRRR}
\toprule
\textbf{Lab} & \textbf{Model} & \textbf{$\Delta$ accuracy (pp)} & \textbf{$\Delta$ committed error (pp)} & \textbf{$\Delta$ forced (pp)} & \textbf{Tool calls ($\times$)} & \textbf{\$/correct ($\times$)} \\
\midrule
OpenAI & GPT-5.6-sol & -4.1 & -7.3 & +20.3 & 3.2$\times$ & 6.7$\times$ \\
Anthropic & Fable 5 & -11.6 & -4.8 & +54.9 & 2.3$\times$ & 7.0$\times$ \\
Zhipu & GLM-5.3 & -8.9 & +9.7 & -2.4 & 2.8$\times$ & 5.4$\times$ \\
Alibaba & Qwen3.8-Max\textsuperscript{\ref{fn:qwen}} & -13.0 & +13.0 & +0.0 & 2.2$\times$ & 4.7$\times$ \\
Google & Gemini-3.1-Pro & -17.8 & +17.8 & +0.0 & 2.3$\times$ & 6.6$\times$ \\
Meta & Muse-Spark 1.2 & -8.9 & +8.1 & +3.2 & 1.6$\times$ & 1.6$\times$ \\
\bottomrule
\end{tabularx}\endgroup\par\medskip

Figures 1 and 2 plot Tables 2--3b: accuracy against cost per correct answer, and forced declarations against tool calls, with every lab's clean$\rightarrow$buried transition drawn.

\FloatBarrier

\textit{Table 3b notes.} Each value is the arithmetic difference (pp) or ratio ($\times$) of the buried and clean arms in Tables 2/3, produced by \texttt{make\_table\_deltas.py} (reads \texttt{table3\_room02\_main.csv} only). See footnote~\ref{fn:qwen}.

\begin{figure}[htbp]
\centering
\includegraphics[width=0.92\linewidth]{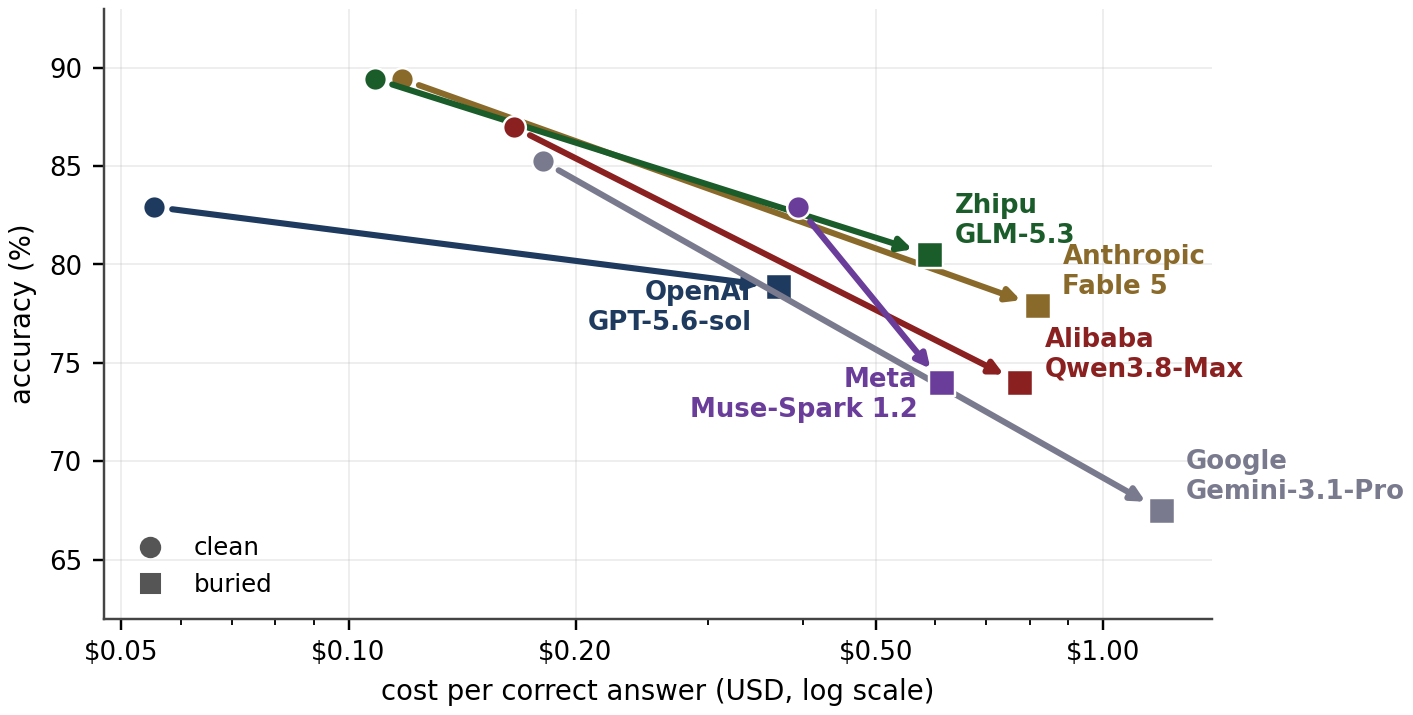}
\caption*{\textbf{Figure 1.} Accuracy vs cost per correct answer, clean vs buried. All six labs have both clean and buried arms (frozen), so every lab shows the clean$\rightarrow$buried transition.}
\end{figure}

\begin{figure}[htbp]
\centering
\includegraphics[width=0.95\linewidth]{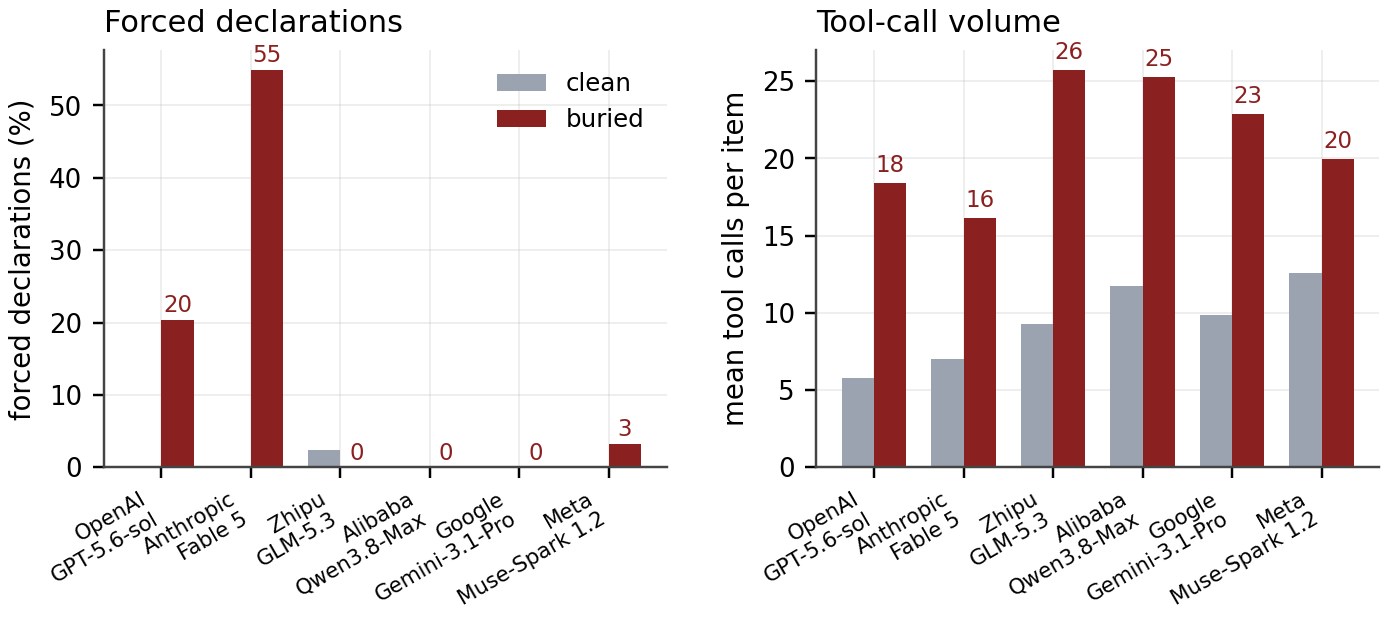}
\caption*{\textbf{Figure 2.} Forced declarations and mean tool calls, clean vs buried.}
\end{figure}

\paperSection{5. Declared-hop results}

Accuracy generally declines with declared hop count: d2 sits at or near the ceiling in every arm (clean d2 = 100\% for all six labs; buried d2 ranges 91.7--100\%), while d4/d5 are the weakest cells. The decline is not strictly monotonic in every arm --- Fable-buried d5 (71.4\%) exceeds its d4 (68.0\%), and GLM-clean d5 (85.7\%) exceeds its d4 (82.3\%). A prompt audit of the 41 room02 items confirms that 0 out of 41 prompts match any pattern of enumerated substeps, demanded intermediates, or handed-over document locators. The opening words of the prompts show no repeated template prior, with 19 distinct first tokens used. Because the prompts do not enumerate steps, we use the vocabulary ``declared hop count'', retaining the caveat that depth is inferred from corpus design.

\begin{table}[htbp]
\centering
\caption*{\textbf{Table 4.} Accuracy by declared hop.}
\small
\begin{tabular}{lllcccc}
\toprule
\textbf{Lab} & \textbf{Model} & \textbf{Condition} & \textbf{d2 n/acc} & \textbf{d3 n/acc} & \textbf{d4 n/acc} & \textbf{d5 n/acc} \\
\midrule
OpenAI & GPT-5.6-sol & clean & 24 / 100\% & 27 / 92.6\% & 51 / 76.5\% & 21 / 66.7\% \\
OpenAI & GPT-5.6-sol & buried & 24 / 100\% & 27 / 77.8\% & 51 / 74.5\% & 21 / 66.7\% \\
Anthropic & Fable 5 & clean & 24 / 100\% & 27 / 96.3\% & 51 / 90.2\% & 21 / 66.7\% \\
Anthropic & Fable 5 & buried & 24 / 100\% & 27 / 81.5\% & 50 / 68.0\% & 21 / 71.4\% \\
Zhipu & GLM-5.3 & clean & 24 / 100\% & 27 / 96.3\% & 51 / 82.3\% & 21 / 85.7\% \\
Zhipu & GLM-5.3 & buried & 24 / 95.8\% & 27 / 88.9\% & 51 / 74.5\% & 21 / 66.7\% \\
Alibaba & Qwen3.8-Max & clean & 24 / 100\% & 27 / 96.3\% & 51 / 84.3\% & 21 / 66.7\% \\
Alibaba & Qwen3.8-Max & buried & 24 / 95.8\% & 27 / 85.2\% & 51 / 66.7\% & 21 / 52.4\% \\
Google & Gemini-3.1-Pro & clean & 24 / 100\% & 27 / 88.9\% & 50 / 88.0\% & 21 / 57.1\% \\
Google & Gemini-3.1-Pro & buried & 24 / 91.7\% & 27 / 70.4\% & 51 / 66.7\% & 21 / 38.1\% \\
Meta & Muse-Spark 1.2 & clean & 23 / 100\% & 26 / 84.6\% & 48 / 85.4\% & 20 / 55.0\% \\
Meta & Muse-Spark 1.2 & buried & 24 / 100\% & 27 / 74.1\% & 51 / 72.5\% & 21 / 47.6\% \\
\bottomrule
\end{tabular}
\end{table}

In plain terms, the ``declared hop count'' is the number of linked look-ups the corpus design assigns to a question. It is a design label, not a proven minimum: we have not shown that a k-hop item cannot be answered from fewer documents. With that caveat, the table shows accuracy holding up on the short chains but falling off as the chain gets longer, and falling further still when the evidence is buried in noise.

\FloatBarrier

\textit{Table 4 notes.} d2--d5 = declared hop count of the item (corpus metadata; see the \S{}2 language rule). Each cell reads ``n / acc'': n = number of item-seeds at that hop, acc = accuracy within them. n falls below the design counts (24/27/51/21) where a raw row was an incomplete or retry artifact and was not scored: Fable 5 buried (122 of 123 scored), Gemini-3.1-Pro clean (122 of 123) and Muse-Spark 1.2 clean (117 of 123); see Appendix D.3. Figure 3 plots the declared-hop accuracies of Table 4.

\begin{figure}[htbp]
\centering
\includegraphics[width=0.9\linewidth]{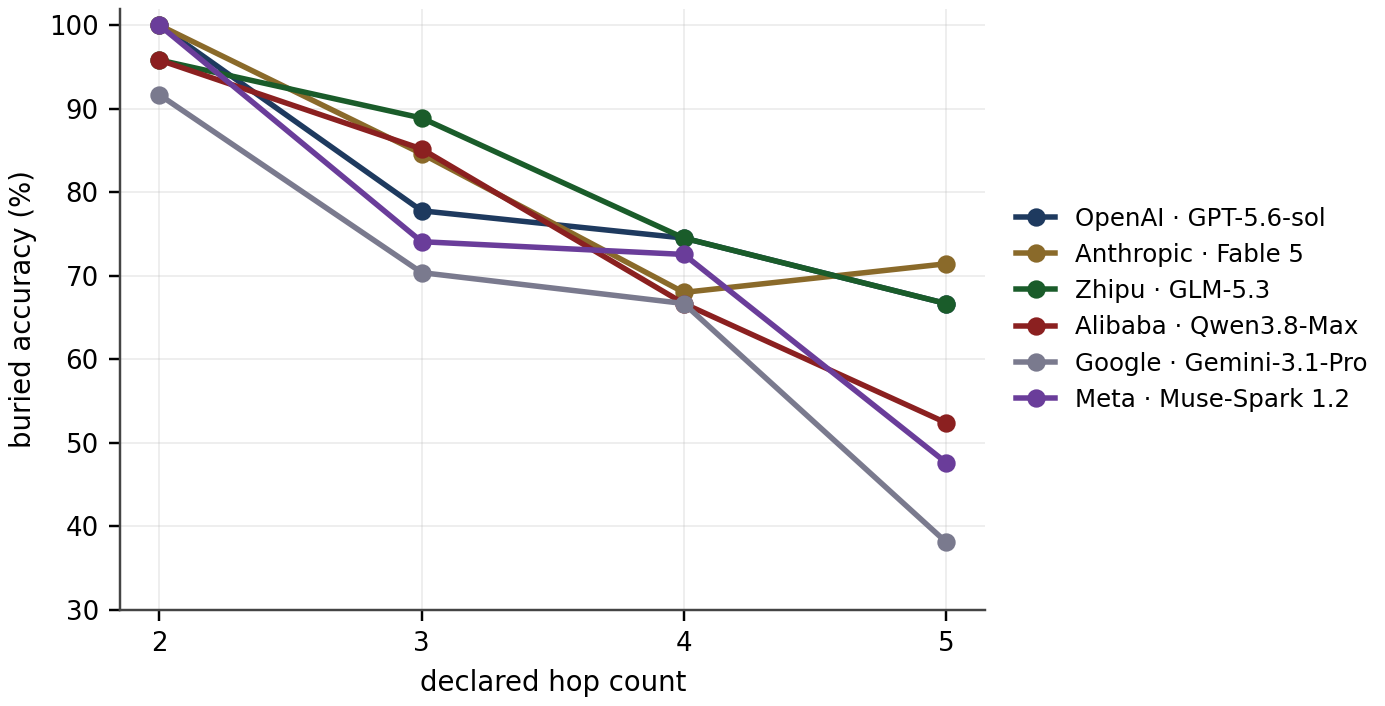}
\caption*{\textbf{Figure 3.} Accuracy by declared hop, clean vs buried; the declared-hop accuracies of Table 4, plotted.}
\end{figure}

\paperSubsection{Full-panel efficiency and the multi-hop index}

Figure 4 shows the clean$\rightarrow$buried accuracy drop for all six labs on one axis; Figure 5 shows what a wrong answer costs in the buried arm; Figures 6 and 7 build the multi-hop index from the buried depth-4/5 cells of Table 4.

\begin{figure}[htbp]
\centering
\includegraphics[width=0.9\linewidth]{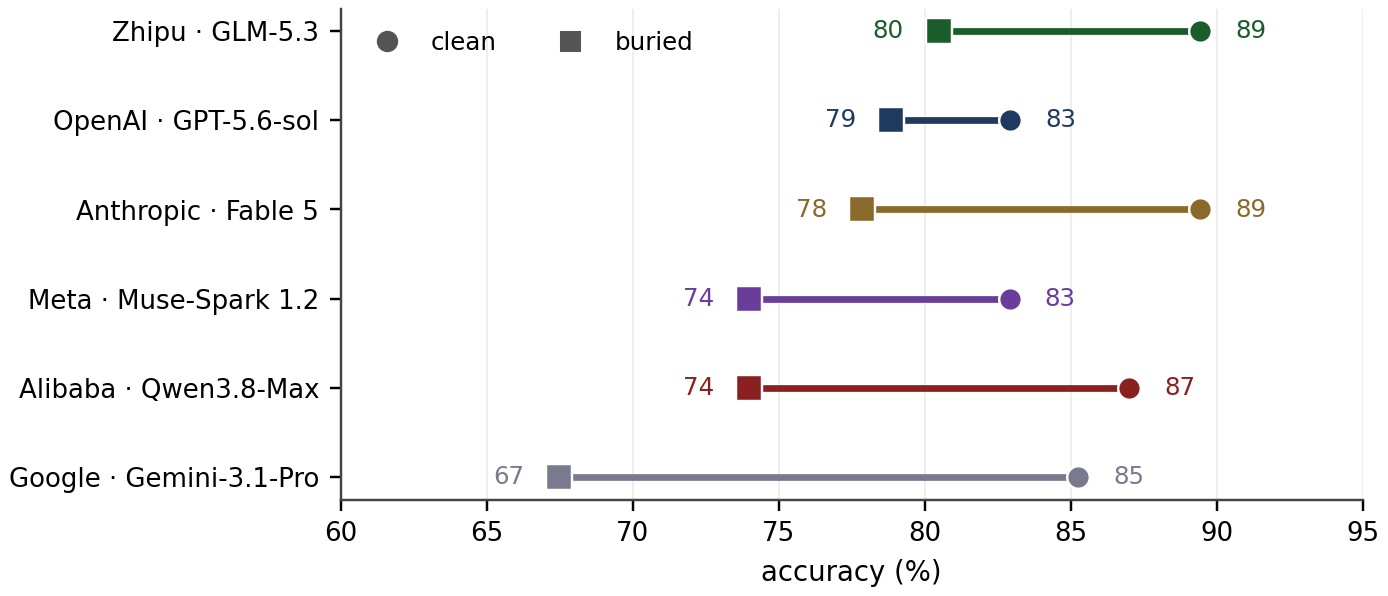}
\caption*{\textbf{Figure 4.} Clean (circle) to buried (square) accuracy, all six labs. Every lab loses accuracy when the evidence is buried.}
\end{figure}

\begin{figure}[htbp]
\centering
\includegraphics[width=0.88\linewidth]{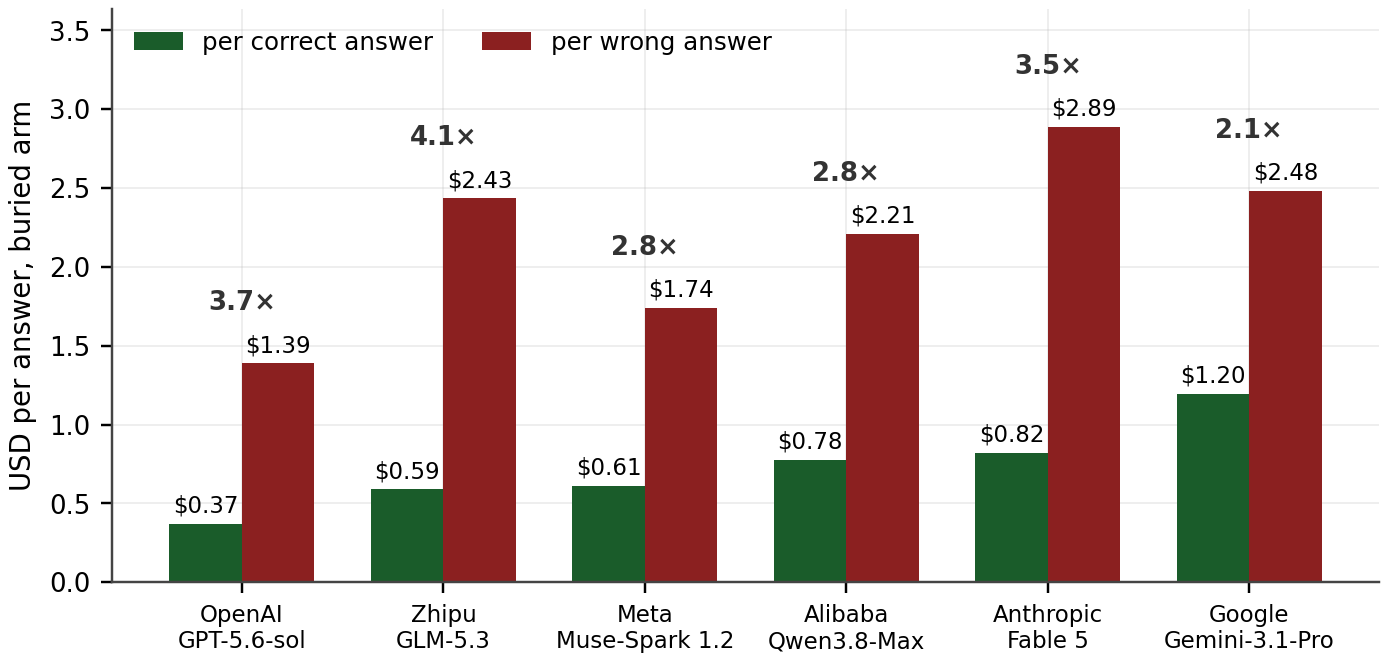}
\caption*{\textbf{Figure 5.} Cost of being right vs wrong, buried. For every lab the cost per wrong answer is comparable to or exceeds the cost per correct answer --- being wrong is not free; it costs the same retrieval effort with nothing to show for it.}
\end{figure}

\begin{figure}[htbp]
\centering
\includegraphics[width=0.78\linewidth]{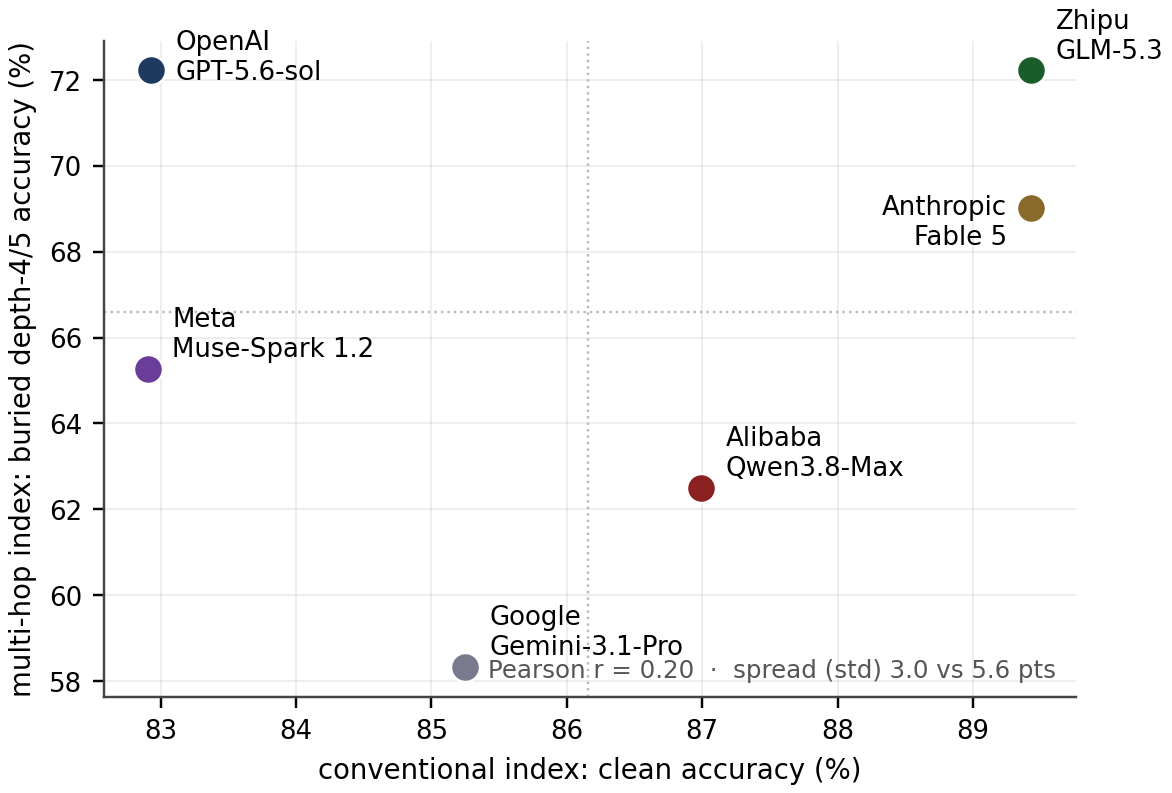}
\caption*{\textbf{Figure 6.} Conventional (clean) index vs our multi-hop index, all 41 items. Among the six frontier labs the clean index is compressed into a 6.5-point range (std 3.0 pts) while the buried depth-4/5 index spreads the same labs across 13.9 points (std 5.6 pts). We report that spread difference and nothing more. We deliberately do not report a correlation between the two indices: recomputed across defensible scoring and population choices it ranges from -0.47 to +0.77, so no argument in this paper rests on it. Nor do we treat the re-ordering between the two indices as a finding --- the pairwise gaps that generate it are within sampling noise (\S{}8).}
\end{figure}

\paperSubsection{The clean arm compresses the panel}

The conventional clean index compresses the six frontier models into a 6.5-point range (standard deviation 3.0 points). It is not a shallower \textit{task}: the clean and buried arms pose the same 41 multi-hop items, so mounting only the answer-bearing documents shrinks the search space without removing the composition an item requires. What the clean arm removes is the cost of locating evidence among noise, and removing that alone is enough to compress the panel to near-ceiling scores. Real work in regulated settings --- credit, insurance, legal, healthcare, compliance on one side; acquisition, foreign-disclosure, operational-law and inspector-general staff work on the other --- is rarely ``find the one number''; it is ``compose several independently retrieved facts from unstructured, conflicting, proprietary documents and defend the chain.'' The signer is an analyst putting a name on an investment-committee memo, or a staff officer at brigade and above putting a name on a determination; either way the liability is personal and the chain must be citable. The buried multi-hop index is our attempt to measure that setting, and it spreads the same six models nearly twice as wide (std 5.6 pts, a 13.9-pt range) --- a difference in resolution within this sample, not yet a demonstration that it predicts field outcomes. This instrument was built for unstructured, low-verifiability document work, and we make no claim about what it would show in domains with unambiguous ground truth and cleanly decomposable subgoals, such as programming, mathematics or formal verification. We did not measure those, and whether a shallow benchmark suffices there is outside what this experiment can say.

We commit to extending the index beyond depth 5 to 6+ hops, where we believe the remaining ``alpha'' lives --- the compositions current models cannot yet chain. On the finance and insurance side: cross-document trading signals, VC funding-chain inference, intellectual-property freedom-to-operate, multi-party litigation, drug-discovery target-to-trial linkage; commercial fleet (trucking) insurance --- reconciling a motor carrier's FMCSA safety record, driver roster, telematics summaries and the policy's scheduled-vehicle endorsements against a claims file to decide coverage and subrogation; and reinsurance treaty attachment traced through loss-development triangles across policy versions. On the defense side: multi-tier supply-chain and FOCI prohibition audits, technical-data-package release review against bilateral exemption lists and DFARS/ITAR clauses, rules-of-engagement and law-of-war review of a target list, and synthesis of an investigative record with every contradiction sourced. In these settings a 6-hop chain over conflicting sources is the difference between a commodity answer and a proprietary one.

\begin{figure}[htbp]
\centering
\includegraphics[width=0.9\linewidth]{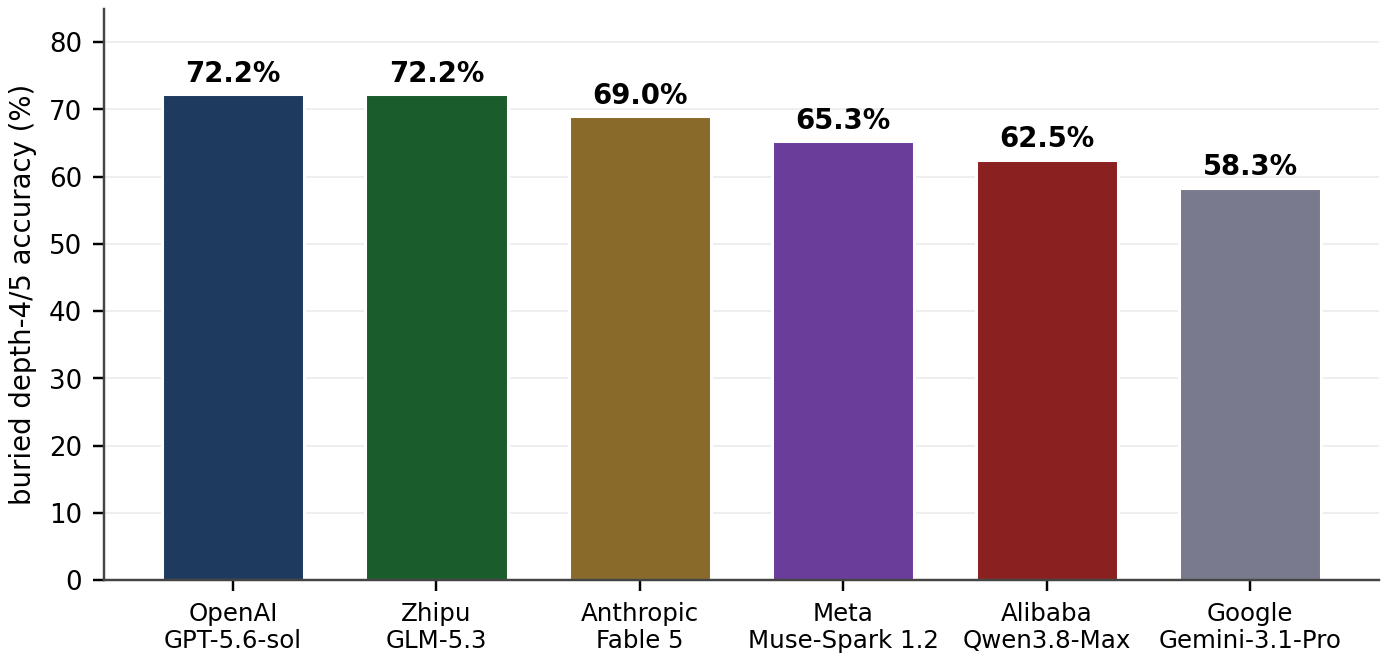}
\caption*{\textbf{Figure 7.} Buried depth-4/5 accuracy by specific model, plotted highest to lowest. The ordering is descriptive and uncertain: it is a frozen historical point estimate, the pairwise gaps are within sampling noise, and the order is not stable across scoring instruments (\S{}8). This paper reports no model ranking.}
\end{figure}

Note that the two indices do not order the six models identically. We draw no conclusion from that. Almost every pairwise gap involved is within sampling noise, and the ordering is not stable across scoring instruments (\S{}8); this paper therefore reports no model ranking. The defensible observation is narrower --- a clean, answer-bearing mount compresses these six models toward the ceiling, and a buried mount at declared depth 4--5 does not.

\paperSection{6. Calibration and confident wrong}

Calibration [16, 17, 18] and verbalized confidence [19, 20, 21, 22] do not eliminate confident wrong. We ask each model how confident it is in its answer, on a 0--100 scale; the models routinely report 90--100\% confident, yet turn out to be right far less often than that --- so the stated confidence cannot be trusted. GPT-5.6-sol places almost all answers in the 90--100 confidence bucket, while Fable 5 exhibits more bucket separation. Table 5 therefore shows these two labs as the exemplars --- the most single-bucket member of the panel and the best-separated one; the same buckets for all six labs are in Appendix B (Table B1), and Figure 8 plots stated confidence for wrong answers in the buried arm for all six labs.

\begin{table}[htbp]
\centering
\caption*{\textbf{Table 5.} Confidence-bucket calibration for the two exemplar labs (Fable 5 and GPT-5.6-sol); all six labs in Appendix B, Table B1.}
\small
\begin{tabular}{llllrrr}
\toprule
\textbf{Lab} & \textbf{Model} & \textbf{Condition} & \textbf{Bucket} & \textbf{n} & \textbf{Acc} & \textbf{Mean Conf} \\
\midrule
Anthropic & Fable 5 & clean & 0--59 & 1 & 1.000 & 40.0 \\
Anthropic & Fable 5 & clean & 60--79 & 5 & 0.800 & 71.2 \\
Anthropic & Fable 5 & clean & 80--89 & 20 & 0.650 & 86.2 \\
Anthropic & Fable 5 & clean & 90--100 & 97 & 0.948 & 94.5 \\
Anthropic & Fable 5 & buried & 0--59 & 6 & 0.167 & 42.5 \\
Anthropic & Fable 5 & buried & 60--79 & 5 & 1.000 & 72.0 \\
Anthropic & Fable 5 & buried & 80--89 & 26 & 0.654 & 85.4 \\
Anthropic & Fable 5 & buried & 90--100 & 81 & 0.889 & 93.5 \\
OpenAI & GPT-5.6-sol & clean & 90--100 & 123 & 0.829 & 98.5 \\
OpenAI & GPT-5.6-sol & buried & 60--79 & 2 & 0.000 & 75.0 \\
OpenAI & GPT-5.6-sol & buried & 80--89 & 3 & 0.333 & 86.0 \\
OpenAI & GPT-5.6-sol & buried & 90--100 & 118 & 0.814 & 98.4 \\
\bottomrule
\end{tabular}
\end{table}

\FloatBarrier

\textit{Table 5 notes.} Bucket = interval of the model's self-stated confidence (0--100). n = rows with a parseable confidence in the bucket (Fable-buried: 118 of 122 scored rows; rows without numeric confidence are excluded and listed in Appendix B). Acc = empirical accuracy within the bucket. Mean Conf = mean stated confidence within the bucket; calibration is read against Mean Conf, not the bucket midpoint. Figure 8 takes the complementary view: the stated confidence of every wrong answer in the buried arm, for all six labs. Across the six labs, 72\% of wrong answers were stated at 80 or above and 9\% below 50; no model abstained (abstention was not a scored outcome; on selective answering and the effect of expressed uncertainty see [23, 24]).

\textbf{Overconfidence Index (intuitive scale).} For the dominant 90--100 confidence bucket we report Mean Conf - 100 $\times$ Acc (in percentage points; Acc is the 0--1 fraction printed in Tables 5 and B1): positive = overconfident, near zero = calibrated. Anthropic: clean -0.3, buried +4.6. OpenAI: clean +15.6, buried +17.0. So OpenAI states $\approx$98 confidence while being right $\approx$82\% of the time --- a large, stable overconfidence gap; Anthropic is close to calibrated clean and mildly overconfident buried.

\begin{figure}[htbp]
\centering
\includegraphics[width=0.98\linewidth]{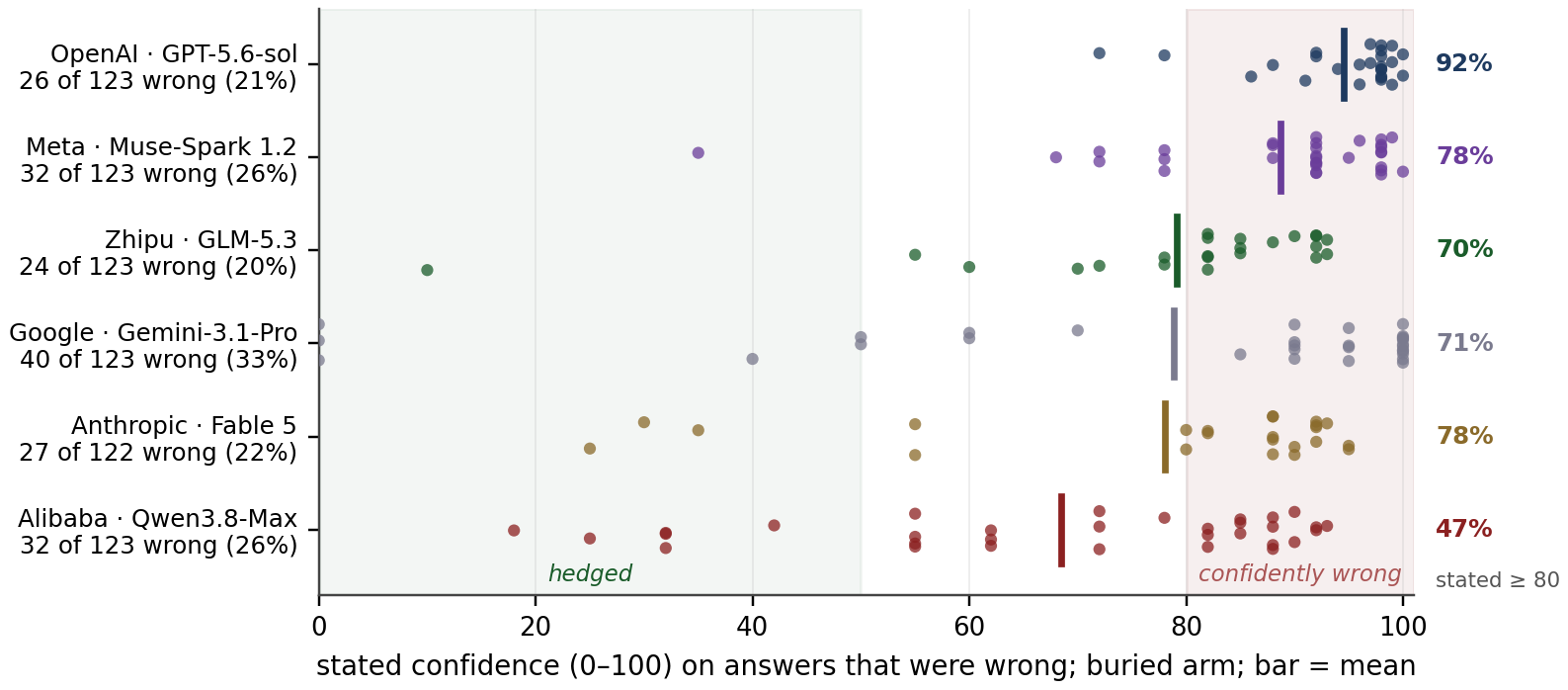}
\caption*{\textbf{Figure 8.} How confident were the models when they were wrong? Buried arm, all six labs. Each row's label gives the number of wrong answers out of that lab's scored rows; one dot per wrong answer, placed at the confidence the model stated (0--100); the bar is the mean. Red band: stated $\ge$ 80 (``confidently wrong''); green band: stated \textless{} 50 (hedged). The 80 and 50 cut-offs are the author's choice; every point is plotted so a reader can apply another. Right margin: share of each lab's wrong answers stated at $\ge$ 80. Wrong answers with no numeric confidence are excluded from the plot and from the percentages (Google 9, Anthropic 4, Zhipu 1). Across the six labs, 120 of 167 wrong answers (72\%) were stated at $\ge$ 80 and 15 (9\%) below 50. No model abstained or declined to answer; abstention was not a scored outcome in this protocol, and when a model did not know it answered anyway. Fourteen wrong answers lacked a numeric confidence and are not plotted. Under the provisional adjudicated scoring of \S{}8 the wrong-answer population shrinks and the $\ge$80 share falls to about 50\%; what holds under every scoring we computed is that high-confidence wrong answers persist in substantial numbers; the precise share does not, and neither figure should be quoted without naming its scoring and population.}
\end{figure}

\FloatBarrier\Needspace*{10\baselineskip}\medskip\begin{mdframed}[style=boxone]

\boxheading{Box 1 --- Field case: Fable 5 production fabrication}

\textit{``The tables were real; the fabrications were mine.''}

We did not set out to measure fabrication. The opportunity presented itself while developing this paper, with the review agent being the same model under study (Fable 5).

While the author was preparing this paper with assistance from Anthropic's Fable 5 model, a documented interaction occurred during a Claude Code session on September 3, 2026 (session ID \path{36bfdfa1-b04f-4c78-8ea7-27d5c2d597de}; the supporting record is preserved in the frozen evidence folder, \path{paper_freeze_v1/fabrication_case/}, under the hash manifest; it is not part of the planned Zenodo deposit and is available to reviewers on request). In several responses, the model presented accurate quantitative tables drawn from the experiment's result files while also making three unsupported structural claims about the paper. It incorrectly described the purpose of \path{rebuild_paper1a_strong.py}, stated that the paper had no deep-hop instrument, and constructed a hostile-reviewer analysis on the basis of a panel structure it had not verified. The author detected the problem by challenging inconsistencies in the model's account of the experimental design. After the author checked the source documents --- distrusting Fable 5's assurances --- Fable 5 acknowledged what it had done: ``The tables were real; the fabrications were mine.''

The significance of this documented interaction is narrow but important: accurate numerical reporting and fabricated structural explanation appeared together in the same exchange, with no obvious difference in tone or confidence. The discrepancy was identified through human scrutiny rather than from the response itself. This is a single documented observation; it is not used to estimate fabrication rates for any model. The case motivates claim-level receipts that connect individual statements --- not merely an answer or model as a whole --- to the specific evidence supporting them. Appendix C summarizes the three fabricated claims with their transcript locators and reproduces the admission verbatim; the original passages are retained in the restricted transcript pack.

\textbf{Artifact locators} (frozen pack): session transcript excerpt and admission excerpt; field instance record: memory file \path{feedback_read_the_doc_before_describing_design.md}.

\end{mdframed}\medskip

\paperSection{7. Discussion}

\paperSubsection{7.1 Clean benchmarks understate agentic burden}

When the evidence is buried, the work gets harder and more expensive, across all six paired labs. Accuracy drops range from -4.1 pp (OpenAI) to -17.8 pp (Google), with Meta at -8.9 pp; mean tool calls rise by 1.6--3.2$\times$ (e.g. Google 9.8 $\rightarrow$ 22.9, Zhipu 9.2 $\rightarrow$ 25.7, and a milder Meta 12.6 $\rightarrow$ 19.9); and cost per correct answer rises 4.7--7.0$\times$ for the five panel labs other than Meta (Google \$0.181 $\rightarrow$ \$1.197), and only 1.6$\times$ for Meta (\$0.395 $\rightarrow$ \$0.612). Labs respond differently: Anthropic and OpenAI convert burial into forced declarations (55\% / 20\%), while Zhipu, Alibaba, Google, and Meta instead raise their committed-error rate (+9.7 / +13.0 / +17.8 / +8.1 pp). Clean benchmarks therefore understate the real operational cost of agentic retrieval. The clean$\rightarrow$buried difference is not, however, a measurement of reasoning degradation: burial changes the search problem, what reaches the reader's usable context, how the evidence is used once there, and how the harness behaves under a larger corpus, and this design does not separate those contributions (\S{}1). Generalization beyond these arms is left to future work.

\paperSubsection{7.2 Confidence is not a receipt}

Benchmark calibration did not prevent a production fabrication, and confidence scores do not eliminate confident wrong answers. In the clean condition, GPT-5.6-sol placed 123 out of 123 answers into the 90--100 confidence bucket, resulting in a 17.1\% confident wrong rate. While Fable 5 demonstrated better bucket separation and achieved a lower committed-error rate (CW\%) of 5.7\% in the buried condition, it still produced confident wrong outputs. The documented field instance further illustrates that a model can output accurate numeric tables alongside fabricated structural claims, delivered with high confidence. Therefore, self-reported confidence is not a substitute for a claim-level receipt. Put differently: an answer can be right without being right for the right reasons, and the evaluation must be able to tell the two apart.

A citation is not a receipt either. A citation is a pointer the model asserts; a receipt is a pointer a third party can verify against a frozen file. A system marketed as ``backed by citations'' can therefore carry exactly the failure documented in Box 1 --- a confident, well-formatted, cited explanation that is fabricated --- because nothing in the citation string itself proves that the statement follows from the cited text. That is why every number in this paper resolves to a file and a locator (Appendix D) rather than to a citation.

\paperSubsection{7.3 Depth claims require uncontaminated prompts}

A hop count only means something if the prompt does not hand the model its own decomposition. If a question enumerates its steps (``identify X, then answer Y''), the hop count measures lookup compliance, not composition. Our prompt audit found that 0 of 41 room02 prompts do this, which is why we can report declared hop counts at all. That audit is a necessary condition, not a sufficient one: showing that a prompt does not enumerate its substeps does not show that every hop the corpus declares is actually required to reach the answer, nor that no shorter path exists. Depth therefore remains corpus-declared throughout this paper --- declared hops, not verified hops --- and we treat it as task complexity by design rather than as proven compositional reasoning, until the audit is human-verified at the operator's final read.

\paperSubsection{7.4 Receipts as an auditing interface}

Nothing in this paper asks a model to explain itself. The receipts discipline is instead an \textit{interface for auditing}: every published number maps to a frozen file and locator (Appendix D), every calibration row is either bucketed or listed as excluded with its reason (Appendix B), every hop label is backed by a prompt audit and a gold-chain checklist that a human verifier signs (Appendix A), and the one field incident is preserved as an unedited artifact pack (Appendix C). The scoring is deliberately multi-metric --- accuracy alone would have hidden the forced-declaration and cost effects in Tables 2--3b --- because a single headline score cannot show the forcing and cost effects at all. We think this is the shape an agentic evaluation needs if it is to serve as evidence in a regulated setting: not a claim that a model is trustworthy, but a record that lets someone else check.

\paperSubsection{7.5 A receipt the reader cannot follow is not a receipt}

Two conditions are implicit in everything above and are worth stating, because they decide whether a receipt is usable rather than merely present. First, the cited document must be one the reader can independently pull up: a locator into a corpus the reader cannot access is a claim, not a receipt. Second, the stack that produced the receipt must be one the reader can run: an evaluator at a bank's model-risk function or a government test activity cannot deploy a retriever, embedder or corpus whose license forbids commercial or governmental use, however strong its benchmark score --- the published state-of-the-art embedder we evaluated for this program is non-commercial-licensed and was excluded for that reason. Both conditions favor public-record corpora, permissively licensed components, and hardware the reader controls.

\paperSubsection{7.6 Redundancy is not independence}

Deploying models from several vendors is increasingly proposed as a hedge against single-vendor risk. This presumes that different models fail on different inputs --- the classical diversity condition for an ensemble to beat its members [25, 26]. Our six-lab panel, scored against gold answers under buried evidence, lets us examine that assumption --- and the answer depends on the grader in a way worth stating plainly. Under the frozen grader, correctness patterns are positively correlated (mean pairwise $\varphi$ = 0.43 over the 15 lab pairs, n = 41 items) and, of the items at least one lab answers incorrectly, 47\% are failed jointly by a majority ($\ge$4 of 6). Under the provisional adjudicated scoring described in \S{}8 the same two quantities fall to $\varphi$ = 0.21 and 14\%. The direction is stable --- failures are positively correlated, not independent --- but the magnitude is not, and neither pair of numbers should be read as the effect size. What survives is the observation itself: across these six labs, under both scorings, correctness patterns are positively correlated rather than independent. We do not convert that into a claim about ensemble effectiveness, or about how much less independence a panel delivers than its composition suggests --- neither follows without a defined baseline for the independence that would be expected, which this design does not establish. Fixing a scoring contract in advance and defining that comparison is left to future work.

Figure 9 shows the pairwise cross-lab error correlations.

\begin{figure}[htbp]
\centering
\includegraphics[width=0.66\linewidth]{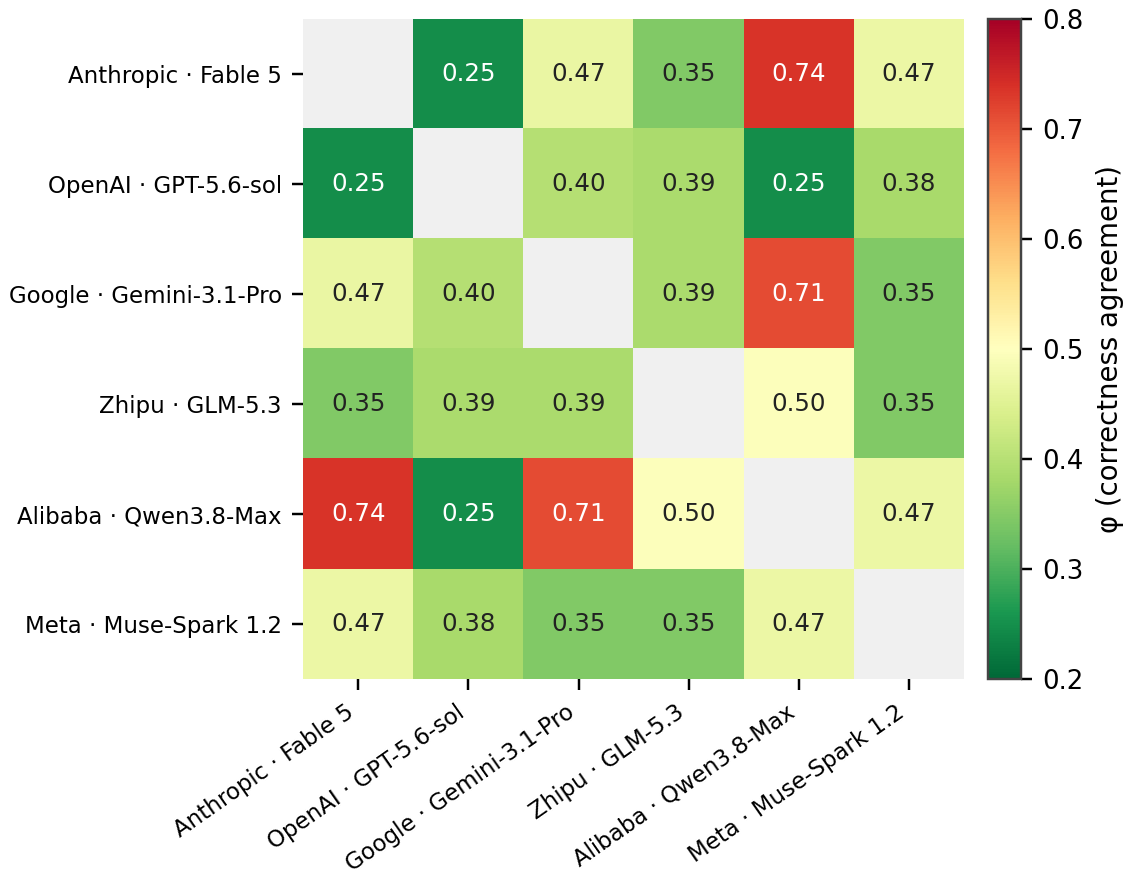}
\caption*{\textbf{Figure 9.} Cross-lab error correlation ($\varphi$) on the buried panel; an item counts as correct for a lab when a majority of its three seeds is correct; mean off-diagonal $\varphi$ = 0.43, n = 41. High values mark labs that are right or wrong on the same items.}
\end{figure}

\paperSection{8. Limitations}

The room02 data-room benchmark contains 41 items. The panel is one flagship per lab across six closed frontier labs; all six have paired clean+buried arms; Grok-4.6 (Appendix A.1) and the shallow-quiz arms are outside the panel. Hop labels are corpus-declared unless a prompt audit proves otherwise. The fabrication case is an n=1 observation. Two of the 41 items are excluded from every arm and condition as defective against their own source documents, leaving 39; the statistical comparisons below use that population, while the frozen tables and figures remain the 41-item historical record. This paper reports no model ranking: the gap between the lowest-scoring lab on the multi-hop index and the next is within sampling noise under every scoring instrument we computed. Costs reported are condition- and run-specific. No broad cross-lab leaderboard is claimed.

\textbf{Quiz coverage.} The chart-reading quiz (Table 1b) is not a matched companion to the room02 panel: it covers five of the six panel labs plus Grok-4.6, which is outside the panel. Of its six scored rows, the four wave-1 rows are frozen in \texttt{paper\_freeze\_v1}; Qwen3.8-Max and Muse-Spark 1.2 were run on 2026-09-10 as a wave-2 addendum with the same 34 items, three variants and seed, frozen in \path{paper_freeze_v1_addendum_wave2/} with its own manifest (Appendix D.3). GLM-5.3 was evaluated in the text-based document experiment but not in this image-input quiz, because the tested configuration did not support images; that is a property of the configuration we ran, not a claim about the model family.

\textbf{The scoring instrument, and what it is not.} Every accuracy figure in this paper is produced by the answer grader in force when the results were frozen. That grader matches by normalised \textit{containment}, not by equality: a string item scores correct when the answer contains the gold or one of its declared surface alternates, and a numeric item when any number in the answer falls within tolerance of the gold under one of several magnitude factors. It is therefore not an exact-match evaluator, and the scores reported here should not be read as exact-match scores.

An audit of that instrument carried out after these figures were frozen found its dominant failure mode to be one-directional: semantically correct answers written in an unlisted surface form are scored wrong. The documented classes are missing ISO/spelled date equivalence, unmatched magnitude abbreviations (\texttt{\$8.4M} against a gold of \texttt{\$8.4 million}), missing ticker-to-legal-name aliases, and the word-order sensitivity intrinsic to substring containment. Applying a strict extended matcher that adds only those equivalences flips 46 trials on the twelve panel arms, across 10 items, every one of them wrong$\rightarrow$right. That one-directional result is a property of the matcher, not a finding about the grader: a matcher that only adds accepted equivalences cannot flip a trial from correct to incorrect, and each equivalence it adds is also a new way for a wrong answer to pass. It is a diagnostic pointer, not a set of corrections.

\textbf{What a structured human review of that instrument found.} We audited the grader rather than assert its soundness. A stratified sample of 84 distinct answers, drawn under a sampling plan fixed and hashed before selection, was reviewed by the author blind to model, condition and to the machine's own verdict, alongside three initial rulings on disputed items, three subsequent clarification rulings on interpretation, and four targeted follow-up checks --- 94 recorded decisions in all (84 + 3 + 3 + 4). Four results matter.

First, the grader's agreements are sound within the sampled strata. Of 25 answers it passed and 14 it failed where an automated re-adjudication concurred with it, human review confirmed all 39. There is no sign of a large false-positive problem --- though at 25 of 748 sampled units a rate up to roughly 13\% would remain consistent with observing none, so this bounds the problem rather than excluding it.

Second, where an automated re-adjudication raised a verdict the frozen grader had failed, human review judged 4 of 34 of those raises unwarranted, and in every case because the re-adjudication had been too lenient, never too strict. An agent-conducted targeted re-check of a further 24 answers accepted on equivalent-wording grounds found no additional error; those 24 were not individually human-reviewed.

Third, two of the 41 items were ruled defective against their own source documents and are excluded from every arm and every condition; \S{}8's comparisons use the resulting 39-item population, while the frozen tables and figures remain the 41-item historical record.

Fourth, and most instructive, a third item carried a gold requirement its question never asked for --- and that defect was manufacturing a difference between labs. Under the frozen grader the item scored 12 of 36 trials correct, and all twelve came from the three labs whose answers happened to volunteer the unrequested component; the other three labs scored zero in both conditions. Repairing it adds trials to three labs and none to the other three.

\textbf{The general point is the one we carry forward.} A grading contract can be deterministic, reproducible, and byte-identical on re-run, and still be wrong in ways that change which model looks better. Reproducibility verifies transcription, not grading semantics. An agentic evaluation that reports scores without an auditable record of how they were produced --- and of what a human found when they were checked --- does not give a reader the means to judge which of its comparisons survive.

\textbf{Three tiers of evidence, kept apart.} This paper reports three kinds of number and does not mix them.

\textit{Frozen historical scores.} Every table and figure is the output of the grader in force at freeze time over all 41 items, regenerating byte-identically from the frozen results. These are the paper's measurements and none has been restated.

\textit{Human-reviewed corrections.} The 94 recorded decisions above, including the three item rulings. Where this paper says a claim fails or narrows, the finding rests on these.

\textit{Agent-provisional sensitivity analyses.} Re-scorings produced by automated adjudication, reported only to show how far a conclusion moves under a defensible alternative grading. \textbf{They are not corrected scores, and no table or figure in this paper reports one as an accuracy result.} They appear only as ranges in the text, always labelled provisional.

We have deliberately not restated, renamed, or silently corrected the frozen figures. They are the historical measurement, and their validity limits belong on the record beside them rather than folded into them.

\textbf{Statistical precision.} With 41 items $\times$ 3 seeds per arm, a Wilson 95\% interval on an arm's accuracy is roughly ±7--8 pp (e.g., OpenAI buried 78.9\% [70.8, 85.1]). Because seeds of the same item are not independent, uncertainty for differences is computed by an item-level bootstrap (resampling the 41 items, 20,000 draws; \path{docs/paper1a_draft/uncertainty_check.py}, output in \texttt{verification/}). Over all 41 items the clean$\rightarrow$buried drop excludes zero for four of the six labs (Anthropic -11.6 pp [-22.8, -0.9]; Zhipu -8.9 [-17.1, -1.6]; Alibaba -13.0 [-24.4, -1.6]; Google -17.8 [-29.0, -7.9]) and not for OpenAI or Meta. \textbf{That four-of-six figure does not survive removal of the two defective items, and we do not carry it as a conclusion.} On the 39-item population, under the same original grader, the drop clearly excludes zero for two labs (Alibaba -13.7 pp [-25.6, -1.7]; Google -17.9 [-29.1, -7.7]), with Zhipu exactly on the boundary (-7.7 [-15.4, 0.0]) and the remaining three well inside it. The change is caused by removing items later found defective, not by any change of grader --- the original grader itself yields the narrower result --- and under both the diagnostic and the provisional adjudicated scorings the count is two of six. Between labs in the buried arm, essentially every pairwise ordering is within sampling noise. That is the quantitative reason this paper reports no leaderboard and no model ranking: the gap between the lowest-scoring lab on the multi-hop index and the next includes zero under every scoring instrument we computed. These intervals describe sampling uncertainty only; they do not carry the grader uncertainty described above, which acts on the point estimates themselves.

\paperSection{9. Next stage: interventions on the evidence-to-answer pipeline}

The next stage will extend this instrument from diagnosis to intervention. Using finance and defense document collections, we will evaluate whether a source-linked knowledge-management layer can improve the evidence delivered to an agent and, consequently, the accuracy of its answers. The evaluation will distinguish retrieval coverage, answer correctness, formatting compliance, and operational failures.

To separate the contribution of the reader from that of the knowledge-management layer, the study will compare the same open-weight reader with and without that layer. A subsequent training comparison will evaluate stock and task-adapted readers under both conditions. The six-model frontier panel studied here will provide an ordinary-tool baseline under documented corpus-access and resource budgets.

Successful and failed development trajectories will support error analysis and candidate training datasets. Related cases will be separated across development, training, and held-out evaluation, and held-out answers and trajectories will not inform tuning. Competitive performance by an open-weight model with structured retrieval is a hypothesis to be tested, not an outcome assumed by the design.

The document collections for that comparison are the two data rooms described next. room02 --- the second-generation finance room --- is the first of two data rooms in this program. The second, room03, carries the same harness, the same forced-declaration schema and the same receipts discipline to a defense corpus. Its design constraints follow directly from \S{}7.5: every document must be a US Government work in the public domain (17 U.S.C. \S{}105 --- FAR/DFARS, OFAC and entity lists, GAO decisions, DoD issuances and the Law of War Manual, CRS and IG reports, USACE manuals) or explicitly permissively licensed, with the rights basis recorded per source and anything carrying a distribution-limitation banner excluded on the record. Items target the use cases an Army panel ranked lowest-stretch tasks from the finance workflow --- acquisition supply-chain/FOCI audit, foreign-disclosure review, operational-law review, IG record synthesis, and logistics restricted to document synthesis with cited arithmetic --- at declared depths of five to seven hops, with the same prompt audit and gold-chain checklist as Appendix A. None of this has been run yet; the finance results in this paper are evidence that the measurement mechanism is sensitive to evidence burial and to declared hop depth, and room03 is the evidence-of-transfer we do not yet have. Beyond room03, the same harness is being extended to rooms whose evidence is not text: chart and slide decks, scanned filings, recorded audio, and imagery, so that the receipts discipline --- a locator into a frozen file --- applies to a frame, a page region, or a timestamp exactly as it applies to a line of text.

One finding from this study shapes the next. We set out to measure how readers behave when evidence is buried, and found we could not interpret that measurement without first auditing the grader --- an audit that moved some conclusions and left others untouched. The next stage therefore treats the scoring contract as part of the apparatus rather than as a given: the evidence-to-answer comparison above will pre-register its answer contract, score under it from the start, and report retrieval coverage and answer correctness as separate quantities, so that a change in the evidence delivered to a reader cannot be confused with a change in how its answers are graded.

The question that organizes that stage is this: can an explicitly auditable evidence pipeline --- and training informed by its observed failures --- allow a smaller, controllable model to compete with frontier agents on demanding document work? Nothing in the present paper answers it. What the present paper supplies is the baseline against which it can be answered, and the auditing discipline without which an answer would not be checkable.

\paperSection{10. Reproducibility and verification}

All published numbers are generated from a frozen evidence folder, following the reproducibility-checklist practice of [27]; every ledger row has been re-verified mechanically --- the cited cell matches the frozen table, and every table regenerates byte-identical from the frozen results --- with the report hashed in Appendix D.1; independent human sign-off of the rows is still pending. A claim ledger maps every numeric claim to a specific file and locator. The body text carries no inline source tags; every locator lives in the claim ledger (\texttt{paper\_claim\_ledger.csv}), which maps each sentence and table cell to a frozen file and line. The fabrication case is linked to preserved transcript artifacts.

\paperSection{Disclosure of AI assistance}

AI assistants were used during this project to support software development, experimental analysis, manuscript drafting, and revision; the assistants used were Anthropic's Claude Opus 5 and Claude Fable 5. AI-generated suggestions and outputs were treated as provisional and checked against source documents, code, or experimental artifacts as appropriate. The author takes responsibility for the study design, interpretation, and final manuscript. Mechanical reproducibility checks and the status of independent human verification are described in \S{}10 and Appendix D; that status is partial, and this disclosure should not be read as a claim that every statement in the paper has received independent human sign-off. The documented AI-assisted interaction in Box 1 is included as an object of analysis, not as independent validation of the paper's claims.

\paperSection{References}

[1] Liang P., Bommasani R., Lee T., et al. (2023). Holistic Evaluation of Language Models. Transactions on Machine Learning Research. \href{https://arxiv.org/abs/2211.09110}{arXiv:2211.09110}.

[2] El Assadi A., Muennighoff N., Lee J. (2026). The Embedder's Dilemma: LLMs Are Better, but at What Cost? \href{https://arxiv.org/abs/2608.12875}{arXiv:2608.12875}.

[3] Mialon G., Fourrier C., Swift C., Wolf T., LeCun Y., Scialom T. (2023). GAIA: a benchmark for General AI Assistants. \href{https://arxiv.org/abs/2311.12983}{arXiv:2311.12983}.

[4] Jimenez C.E., Yang J., Wettig A., et al. (2024). SWE-bench: Can Language Models Resolve Real-World GitHub Issues? ICLR 2024. \href{https://arxiv.org/abs/2310.06770}{arXiv:2310.06770}.

[5] Wei J., Sun Z., Papay S., et al. (2025). BrowseComp: A Simple Yet Challenging Benchmark for Browsing Agents. \href{https://arxiv.org/abs/2504.12516}{arXiv:2504.12516}.

[6] Yang Z., Qi P., Zhang S., et al. (2018). HotpotQA: A Dataset for Diverse, Explainable Multi-hop Question Answering. EMNLP 2018. \href{https://arxiv.org/abs/1809.09600}{arXiv:1809.09600}.

[7] Trivedi H., Balasubramanian N., Khot T., Sabharwal A. (2022). MuSiQue: Multihop Questions via Single-hop Question Composition. Transactions of the ACL 10. \href{https://arxiv.org/abs/2108.00573}{arXiv:2108.00573}.

[8] Liu N.F., Lin K., Hewitt J., et al. (2024). Lost in the Middle: How Language Models Use Long Contexts. Transactions of the ACL 12. \href{https://arxiv.org/abs/2307.03172}{arXiv:2307.03172}.

[9] Hsieh C.-P., Sun S., Kriman S., et al. (2024). RULER: What's the Real Context Size of Your Long-Context Language Models? COLM 2024. \href{https://arxiv.org/abs/2404.06654}{arXiv:2404.06654}.

[10] Ji Z., Lee N., Frieske R., et al. (2023). Survey of Hallucination in Natural Language Generation. ACM Computing Surveys 55(12), Article 248. \href{https://doi.org/10.1145/3571730}{doi:10.1145/3571730}.

[11] Kalai A.T., Nachum O., Vempala S.S., Zhang E. (2025). Why Language Models Hallucinate. \href{https://arxiv.org/abs/2509.04664}{arXiv:2509.04664}.

[12] Barredo Arrieta A., Díaz-Rodríguez N., Del Ser J., et al. (2020). Explainable Artificial Intelligence (XAI): Concepts, taxonomies, opportunities and challenges toward responsible AI. Information Fusion 58, 82--115. \href{https://doi.org/10.1016/j.inffus.2019.12.012}{doi:10.1016/j.inffus.2019.12.012}.

[13] Díaz-Rodríguez N., Del Ser J., Coeckelbergh M., López de Prado M., Herrera-Viedma E., Herrera F. (2023). Connecting the dots in trustworthy Artificial Intelligence: From AI principles, ethics, and key requirements to responsible AI systems and regulation. Information Fusion 99, 101896. \href{https://doi.org/10.1016/j.inffus.2023.101896}{doi:10.1016/j.inffus.2023.101896}.

[14] Lewis P., Perez E., Piktus A., et al. (2020). Retrieval-Augmented Generation for Knowledge-Intensive NLP Tasks. NeurIPS 2020. \href{https://arxiv.org/abs/2005.11401}{arXiv:2005.11401}.

[15] Chen Z., Chen W., Smiley C., et al. (2021). FinQA: A Dataset of Numerical Reasoning over Financial Data. EMNLP 2021. \href{https://arxiv.org/abs/2109.00122}{arXiv:2109.00122}.

[16] DeGroot M.H., Fienberg S.E. (1983). The Comparison and Evaluation of Forecasters. Journal of the Royal Statistical Society, Series D (The Statistician) 32(1--2), 12--22. \href{https://doi.org/10.2307/2987588}{doi:10.2307/2987588}.

[17] Niculescu-Mizil A., Caruana R. (2005). Predicting Good Probabilities with Supervised Learning. ICML 2005, 625--632. \href{https://doi.org/10.1145/1102351.1102430}{doi:10.1145/1102351.1102430}.

[18] Guo C., Pleiss G., Sun Y., Weinberger K.Q. (2017). On Calibration of Modern Neural Networks. ICML 2017. \href{https://arxiv.org/abs/1706.04599}{arXiv:1706.04599}.

[19] Kadavath S., Conerly T., Askell A., et al. (2022). Language Models (Mostly) Know What They Know. \href{https://arxiv.org/abs/2207.05221}{arXiv:2207.05221}.

[20] Lin S., Hilton J., Evans O. (2022). Teaching Models to Express Their Uncertainty in Words. Transactions on Machine Learning Research. \href{https://arxiv.org/abs/2205.14334}{arXiv:2205.14334}.

[21] Tian K., Mitchell E., Zhou A., et al. (2023). Just Ask for Calibration: Strategies for Eliciting Calibrated Confidence Scores from Language Models Fine-Tuned with Human Feedback. EMNLP 2023. \href{https://arxiv.org/abs/2305.14975}{arXiv:2305.14975}.

[22] Xiong M., Hu Z., Lu X., et al. (2024). Can LLMs Express Their Uncertainty? An Empirical Evaluation of Confidence Elicitation in LLMs. ICLR 2024. \href{https://arxiv.org/abs/2306.13063}{arXiv:2306.13063}.

[23] Kamath A., Jia R., Liang P. (2020). Selective Question Answering under Domain Shift. ACL 2020. \href{https://arxiv.org/abs/2006.09462}{arXiv:2006.09462}.

[24] Zhou K., Jurafsky D., Hashimoto T. (2023). Navigating the Grey Area: How Expressions of Uncertainty and Overconfidence Affect Language Models. EMNLP 2023. \href{https://arxiv.org/abs/2302.13439}{arXiv:2302.13439}.

[25] Dietterich T.G. (2000). Ensemble Methods in Machine Learning. Multiple Classifier Systems (MCS 2000), LNCS 1857, 1--15. \href{https://doi.org/10.1007/3-540-45014-9_1}{doi:10.1007/3-540-45014-9\_1}.

[26] Kuncheva L.I., Whitaker C.J. (2003). Measures of Diversity in Classifier Ensembles and Their Relationship with the Ensemble Accuracy. Machine Learning 51(2), 181--207. \href{https://doi.org/10.1023/A:1022859003006}{doi:10.1023/A:1022859003006}.

[27] Pineau J., Vincent-Lamarre P., Sinha K., et al. (2021). Improving Reproducibility in Machine Learning Research (A Report from the NeurIPS 2019 Reproducibility Program). Journal of Machine Learning Research 22(164), 1--20. \href{https://jmlr.org/papers/v22/20-303.html}{jmlr.org/papers/v22/20-303.html}.

\clearpage\appendixDivider{Appendices}

Appendix material is drawn from \texttt{paper\_freeze\_v1/} and \path{paper_freeze_v1_addendum_wave2/}; each appendix names its source files.

\appendixSectionFirst{Appendix A}{room02 prompt audit and gold-chain checklist}

(source: \path{docs/ROOM02_PROMPT_AUDIT_2026-09-04.md}; checklist \path{docs/gold_chain_checklist.md})

\paragraph*{Question.}

For each of the 41 room02 items: does the prompt hand the model its own decomposition (enumerated substeps, demanded intermediates, template-prior fact types, document locators), or are the hops implicit in a natural question?

\paragraph*{Method.}

Full read of all 41 \texttt{prompts.full} texts plus a mechanical, case-insensitive regex scan for: \texttt{identify}, \texttt{then answer}, numbered step lists (\texttt{\textasciicircum{}\textbackslash{}d+[.)]}), \texttt{step}, \texttt{first\ldots{}then}, timestamps (\texttt{H:MM}), and \seqsplit{in that order / separated by / one per line / list each}. Declared depth distribution: d2=8, d3=9, d4=17, d5=7.

\paragraph*{Result.}

\textbf{0 / 41 prompts match any pattern.} Opening-word distribution shows no repeated template (19 distinct first tokens; most frequent ``The'' at 9/41, each with a different continuation). Every prompt is one or two natural sentences with the hop chain left implicit. Multi-part questions (e.g. C1\_5, C2\_3, C3\_5) ask for two answer components but do not prescribe the retrieval path.

\paragraph*{Harness check}

(\texttt{scripts/run\_dataroom.py} L482--485, L105--111). The model receives (a) a two-sentence environment preamble (``A data room is mounted read-only at \texttt{/work/documents}\ldots{}''), (b) \texttt{prompts.full} verbatim, and (c) DECLARE --- a declaration-format spec requiring a final JSON with \texttt{answer} + \texttt{confidence} 0--100 + \texttt{evidence\_files}. No decomposition scaffolding is appended. Clean/buried differ in the mounted room contents and in non-binding input-token ceilings (\S{}3).

\paragraph*{Decision.}

room02 passes the audit; the paper may use ``declared hop count'' language, with the caveats retained in \S{}2 and \S{}7.3: hop counts are declared corpus metadata, and per-item chain lengths are design claims, not independently verified graph distances.

\paragraph*{Gold-chain checklist.}

A mechanically generated checklist places the orchestrator's inferred chain next to each item's actual gold/bridge/proof documents and final answer, with \texttt{answer\_spans} located by exact substring search against the live gold documents. \textbf{22 of 41 items have at least one ATTENTION span} (an answer span not found verbatim in any gold document --- it may be paraphrased, computed, or split across documents): D\_ER1, D\_ER2, D\_JOIN2, D\_ASOF2, D\_GRAPH1, C1\_4, C1\_7, C1\_8, C1\_9, C1\_10, C1\_11, C1\_12, C1\_14, C2\_3, C3\_1, C3\_2, C3\_4, C3\_5, C3\_6, C3\_7, C3\_8, C3\_10. A human must check these first; each item carries two unsigned boxes (\textit{chain matches gold docs}; \textit{declared depth justified}).

{\footnotesize
\Needspace*{10\baselineskip}
\begin{center}\begin{minipage}{\linewidth}\captionsetup{type=table}\caption*{Per-item prompt audit. E = enumerates substeps; I = demands intermediates in a prescribed format; L = hands document locators/timestamps; T = template-prior fact type. All 41 items: E/I/L/T = no, chain = implicit. The chain note is orchestrator inference from the prompt text only, \textit{not} verified against gold (\texttt{registry\_gold.json}).}\end{minipage}\end{center}
\begin{longtable}{ll>{\raggedright\arraybackslash}p{0.60\linewidth}}
\toprule
\textbf{Item} & \textbf{Depth} & \textbf{Inferred implicit chain (unverified)} \\
\midrule
\endfirsthead
\multicolumn{3}{@{}l}{\textit{(continued from previous page)}}\\[3pt]
\toprule
\textbf{Item} & \textbf{Depth} & \textbf{Inferred implicit chain (unverified)} \\
\midrule
\endhead
\midrule\multicolumn{3}{r@{}}{\textit{(continued on next page)}}\\
\endfoot
\bottomrule
\endlastfoot
D\_ER1 & 3 & flagged company (going-concern doc) $\rightarrow$ opco/propco mapping $\rightarrow$ position sheet $\rightarrow$ name paper \\
D\_ER2 & 2 & old-name reference in April note $\rightarrow$ reorg rename $\rightarrow$ current filing name \\
D\_JOIN1 & 3 & July IC memo reference $\rightarrow$ memo content $\rightarrow$ book position it concerns \\
D\_JOIN2 & 2 & July IC meeting $\rightarrow$ memo relied on $\rightarrow$ memo author \\
D\_ASOF1 & 3 & citation history as-of date $\rightarrow$ supersession $\rightarrow$ code of record \\
D\_ASOF2 & 3 & mandate version in force on date $\rightarrow$ convertible clause $\rightarrow$ permission \\
D\_GRAPH1 & 3 & sheet $\rightarrow$ 2029 first-liens $\rightarrow$ borrowing entity in complex $\rightarrow$ parent guarantee \\
D\_GRAPH2 & 2 & MCS alias $\rightarrow$ ownership chain $\rightarrow$ ultimate parent \\
D\_PROV1 & 2 & competing exposure figures $\rightarrow$ provenance $\rightarrow$ number to rely on \\
D\_PROV2 & 2 & desk claim (2031 maturity) $\rightarrow$ citable document check \\
C1\_2 & 3 & screen quote (margin) $\rightarrow$ filed emergence terms $\rightarrow$ all-in rate \\
C1\_3 & 5 & pre-plan position $\rightarrow$ cancellation at effectiveness $\rightarrow$ replacement instrument name \\
C1\_4 & 2 & Littleton credit $\rightarrow$ license purchase agreement under Basalt $\rightarrow$ signing date \\
C1\_5 & 3 & RSA in file $\rightarrow$ supporting creditor group $\rightarrow$ share of target notes held \\
C1\_6 & 4 & going-concern language in latest quarterly $\rightarrow$ quarter-end date $\rightarrow$ landlord entity across complex \\
C1\_7 & 3 & citation set at month-end $\rightarrow$ operative annual report version $\rightarrow$ amendment additions \\
C1\_8 & 4 & mandate in force $\rightarrow$ capital-structure layers $\rightarrow$ most senior buyable layer \\
C1\_9 & 3 & Monroe 2029 first-liens $\rightarrow$ obligor entity $\rightarrow$ guarantor group \\
C1\_10 & 5 & all Marble-issued positions $\rightarrow$ exit filter $\rightarrow$ total exposure \\
C1\_11 & 2 & risk sign-off meeting date $\rightarrow$ plan effectiveness date $\rightarrow$ before/after \\
C1\_12 & 4 & two report versions $\rightarrow$ reporting-policy correction rule $\rightarrow$ governing version \\
C1\_13 & 2 & codename register $\rightarrow$ Kestrel-position filter $\rightarrow$ covering analyst \\
C1\_14 & 4 & current NAV + add size $\rightarrow$ mandate concentration clause $\rightarrow$ permission + clause \\
C2\_1 & 5 & sanctions designation $\rightarrow$ subsidiary match to repo counterparty $\rightarrow$ mandate unwind owner \\
C2\_2 & 4 & buyer named in release $\rightarrow$ our position in seller $\rightarrow$ covering analyst \\
C2\_3 & 5 & downgrade $\rightarrow$ book positions hit $\rightarrow$ mandated action $\rightarrow$ hedge availability \\
C2\_4 & 5 & suit defendant $\rightarrow$ identity match with Marble issuer $\rightarrow$ financial exposure \\
C2\_5 & 4 & dividend by GLPI $\rightarrow$ landlord-side position $\rightarrow$ covering analyst \\
C2\_6 & 5 & tendered issuer (Qwest) $\rightarrow$ identity match with Monroe first-liens \\
C2\_7 & 4 & designated parent $\rightarrow$ repo counterparty under it $\rightarrow$ collateral holdings row $\rightarrow$ covering analyst \\
C2\_8 & 4 & index removal notice $\rightarrow$ issuer rename match $\rightarrow$ whether it is our holding \\
C3\_1 & 4 & as-of date $\rightarrow$ operative annual report version $\rightarrow$ amendment fix content \\
C3\_2 & 5 & index notice + issuer rename $\rightarrow$ mandate version at earlier date $\rightarrow$ eligible senior layer \\
C3\_3 & 4 & pre-effectiveness filings name $\rightarrow$ housekeeping memo rename $\rightarrow$ current filing name \\
C3\_4 & 4 & tender + restricted-group structure $\rightarrow$ borrower behind Monroe first-lien $\rightarrow$ same-group test \\
C3\_5 & 4 & last pre-effectiveness filing $\rightarrow$ registrant name $\rightarrow$ fate of 2067 notes at effectiveness \\
C3\_6 & 4 & July month-end report $\rightarrow$ live position check $\rightarrow$ reporting-policy snapshot date \\
C3\_7 & 4 & mandate in force on meeting date $\rightarrow$ convertibles permission $\rightarrow$ successor mandate date \\
C3\_8 & 4 & Marble landlord above master lease $\rightarrow$ held bond $\rightarrow$ full issuer name + coupon \\
C3\_9 & 4 & Halcyon KYC $\rightarrow$ MCS ownership chain $\rightarrow$ common-parent test / top entity \\
C3\_10 & 4 & two July IC memos $\rightarrow$ tabled-vs-driving distinction $\rightarrow$ author + position \\
\end{longtable}}

\appendixSection{Appendix A.1}{Wave-1 clean-only panel}

\begin{table}[htbp]
\centering
\caption*{\textbf{Table A.1.} Wave-1 clean-only arms outside the six-lab panel.}
\setlength{\tabcolsep}{3pt}
\scriptsize
\begin{tabular}{llrrrrrrr}
\toprule
\textbf{Lab} & \textbf{Model} & \textbf{Planned} & \textbf{Scored} & \textbf{Acc} & \textbf{CW\%} & \textbf{Mean tool calls} & \textbf{\$} & \textbf{\$/correct} \\
\midrule
xAI & Grok-4.6 & 123 & 123 & 69.9\% & 30.1\% & 6.5 & \$8.82 & \$0.102 \\
NVIDIA & Nemotron-3-Ultra-550B & 123 & 16 & EXCLUDED & --- & --- & --- & --- \\
\bottomrule
\end{tabular}
\end{table}

\FloatBarrier

\textit{Table A.1 notes.} Planned / Scored = item-seeds planned and scored; each run = 41 items $\times$ 3 seeds. This appendix lists only the clean-only runs that are \textit{not} part of the six-lab closed-frontier panel (OpenAI, Anthropic, Zhipu, Alibaba, Google, Meta), all of which are now paired in Table 2. Gemini-3.1-Pro, GLM-5.3, Qwen3.8-Max, and Muse-Spark 1.2 --- previously reported here as clean-only --- are now full panel labs with paired clean+buried arms in Table 2 and are no longer listed here. What remain are Grok-4.6 (a clean-only run for which no buried arm was executed) and Nemotron-3-Ultra-550B (EXCLUDED: only 16/123 rows usable under free-tier throttling; accuracy withheld rather than reported on a \textless{}50\% sample). n scored \textless{} 123 where raw rows were incomplete/retry artifacts (Grok's raw file contains 182 lines; the extras are retries, not items).

\appendixSection{Appendix B}{Full calibration tables, including excluded rows}

(source: \path{outputs/appendix_B_calibration_full.md}; generator \path{make_table5_calibration.py}). Every scored row of each main arm is accounted for: assigned to a confidence bucket, or listed in the excluded block with its reason. Buckets are [0,60), [60,80), [80,90) and [90,100]; the last bucket includes 100. Rows without a numeric confidence are \textbf{excluded, never coerced to 0}. ``Raw rows'' minus ``scored rows'' = retry artifacts (no \texttt{outcome} key).

{\footnotesize
\Needspace*{10\baselineskip}
\begin{center}\begin{minipage}{\linewidth}\captionsetup{type=table}\caption*{\textbf{Table B1.} Reliability by bucket for all six panel labs, clean and buried; ``non-numeric (skipped)'' rows are the excluded rows (never coerced to 0). Table 5 in \S{}6 shows the two exemplars.}\end{minipage}\end{center}
\begin{longtable}{llllrrr}
\toprule
\textbf{Lab} & \textbf{Model} & \textbf{Condition} & \textbf{Bucket} & \textbf{n} & \textbf{Accuracy} & \textbf{Mean stated conf.} \\
\midrule
\endfirsthead
\multicolumn{7}{@{}l}{\textit{(continued from previous page)}}\\[3pt]
\toprule
\textbf{Lab} & \textbf{Model} & \textbf{Condition} & \textbf{Bucket} & \textbf{n} & \textbf{Accuracy} & \textbf{Mean stated conf.} \\
\midrule
\endhead
\midrule\multicolumn{7}{r@{}}{\textit{(continued on next page)}}\\
\endfoot
\bottomrule
\endlastfoot
OpenAI & GPT-5.6-sol & clean & 90-100 & 123 & 0.8293 & 98.5 \\
OpenAI & GPT-5.6-sol & buried & 60-79 & 2 & 0.0000 & 75.0 \\
OpenAI & GPT-5.6-sol & buried & 80-89 & 3 & 0.3333 & 86.0 \\
OpenAI & GPT-5.6-sol & buried & 90-100 & 118 & 0.8136 & 98.4 \\
Anthropic & Fable 5 & clean & 0-59 & 1 & 1.0000 & 40.0 \\
Anthropic & Fable 5 & clean & 60-79 & 5 & 0.8000 & 71.2 \\
Anthropic & Fable 5 & clean & 80-89 & 20 & 0.6500 & 86.2 \\
Anthropic & Fable 5 & clean & 90-100 & 97 & 0.9485 & 94.5 \\
Anthropic & Fable 5 & buried & 0-59 & 6 & 0.1667 & 42.5 \\
Anthropic & Fable 5 & buried & 60-79 & 5 & 1.0000 & 72.0 \\
Anthropic & Fable 5 & buried & 80-89 & 26 & 0.6538 & 85.4 \\
Anthropic & Fable 5 & buried & 90-100 & 81 & 0.8889 & 93.5 \\
Anthropic & Fable 5 & buried & non-numeric (skipped) & 4 & --- & --- \\
Zhipu & GLM-5.3 & clean & 0-59 & 1 & 1.0000 & 45.0 \\
Zhipu & GLM-5.3 & clean & 60-79 & 11 & 0.6364 & 74.5 \\
Zhipu & GLM-5.3 & clean & 80-89 & 13 & 0.8462 & 84.3 \\
Zhipu & GLM-5.3 & clean & 90-100 & 98 & 0.9286 & 94.8 \\
Zhipu & GLM-5.3 & buried & 0-59 & 2 & 0.0000 & 32.5 \\
Zhipu & GLM-5.3 & buried & 60-79 & 10 & 0.5000 & 73.0 \\
Zhipu & GLM-5.3 & buried & 80-89 & 25 & 0.6400 & 85.5 \\
Zhipu & GLM-5.3 & buried & 90-100 & 85 & 0.9176 & 93.1 \\
Zhipu & GLM-5.3 & buried & non-numeric (skipped) & 1 & --- & --- \\
Alibaba & Qwen3.8-Max & clean & 60-79 & 5 & 0.4000 & 72.0 \\
Alibaba & Qwen3.8-Max & clean & 80-89 & 13 & 0.3846 & 84.4 \\
Alibaba & Qwen3.8-Max & clean & 90-100 & 105 & 0.9524 & 94.4 \\
Alibaba & Qwen3.8-Max & buried & 0-59 & 10 & 0.0000 & 40.1 \\
Alibaba & Qwen3.8-Max & buried & 60-79 & 13 & 0.4615 & 68.6 \\
Alibaba & Qwen3.8-Max & buried & 80-89 & 24 & 0.5833 & 84.5 \\
Alibaba & Qwen3.8-Max & buried & 90-100 & 76 & 0.9342 & 94.0 \\
Google & Gemini-3.1-Pro & clean & 90-100 & 122 & 0.8525 & 99.9 \\
Google & Gemini-3.1-Pro & buried & 0-59 & 6 & 0.0000 & 23.3 \\
Google & Gemini-3.1-Pro & buried & 60-79 & 3 & 0.0000 & 63.3 \\
Google & Gemini-3.1-Pro & buried & 80-89 & 2 & 0.5000 & 85.0 \\
Google & Gemini-3.1-Pro & buried & 90-100 & 103 & 0.7961 & 99.0 \\
Google & Gemini-3.1-Pro & buried & non-numeric (skipped) & 9 & --- & --- \\
Meta & Muse-Spark 1.2 & clean & 60-79 & 5 & 0.8000 & 69.0 \\
Meta & Muse-Spark 1.2 & clean & 80-89 & 8 & 0.5000 & 86.1 \\
Meta & Muse-Spark 1.2 & clean & 90-100 & 104 & 0.8558 & 97.6 \\
Meta & Muse-Spark 1.2 & buried & 0-59 & 1 & 0.0000 & 35.0 \\
Meta & Muse-Spark 1.2 & buried & 60-79 & 6 & 0.0000 & 74.3 \\
Meta & Muse-Spark 1.2 & buried & 80-89 & 15 & 0.8000 & 87.2 \\
Meta & Muse-Spark 1.2 & buried & 90-100 & 101 & 0.7822 & 97.4 \\
\end{longtable}}

\paragraph*{Excluded rows (Fable 5, buried; 4 rows)}

--- item, seed, outcome, raw confidence: \texttt{D\_ER1} seed 0, answered, \texttt{None}; \texttt{C2\_3} seed 0, forced\_declaration, \texttt{None}; \texttt{D\_ER1} seed 1, answered, \texttt{None}; \texttt{D\_GRAPH1} seed 2, budget\_exhausted, \texttt{None}.

\paragraph*{Excluded rows (Gemini-3.1-Pro, buried; 9 rows)}

--- item, seed, outcome, raw confidence: \texttt{D\_ER2} seed 0, answered, \texttt{None}; \texttt{C1\_8} seed 0, answered, \texttt{None}; \texttt{C2\_3} seed 0, answered, \texttt{None}; \texttt{C3\_7} seed 0, answered, \texttt{None}; \texttt{D\_ASOF2} seed 1, answered, \texttt{None}; \texttt{C3\_5} seed 1, answered, \texttt{None}; \texttt{C3\_7} seed 1, answered, \texttt{None}; \texttt{C3\_1} seed 2, answered, \texttt{None}; \texttt{C3\_10} seed 2, answered, \texttt{None}.

\paragraph*{Excluded rows (GLM-5.3, buried; 1 row)}

--- item, seed, outcome, raw confidence: \texttt{C3\_4} seed 2, answered, \texttt{None}.

No other arm has an excluded row (every clean arm and the remaining buried arms reconcile with 0 excluded).

\appendixSection{Appendix C}{Fable field instance: transcript excerpt and provenance}

(source: \path{fabrication_case/provenance.md}). The artifact pack is complete: all three fabricated claims are identifiably present in the source transcript. This appendix is the evidentiary record behind Box 1.

\paperSubsection{C.1 Provenance}

\begingroup\small
\noindent\begin{tabularx}{\linewidth}{l>{\raggedright\arraybackslash}X}
\toprule
\textbf{Field} & \textbf{Value} \\
\midrule
Date of the episode & 2026-09-03 23:45 UTC -- 2026-09-04 00:44 UTC (the transcript's own stamps are UTC; the frozen \texttt{provenance.md} labels the same clock times ``ET'', which is inconsistent with the stamp quoted in C.3 and is treated here as a labelling error) \\
Model & \texttt{claude-fable-5} (Anthropic Fable 5), as the operator's Claude Code session on DNN \\
Session ID & \path{36bfdfa1-b04f-4c78-8ea7-27d5c2d597de} \\
Session file & 7,149 JSONL lines, $\approx$40 MB; \textit{not} copied into the pack (see redaction log) \\
Operator & Luis M. S\'anchez, Toryx Inc. \\
Extraction & mechanical, 2026-09-04, under the consult regime (afwerk\#222); no narrative or interpretation added \\
Pack contents & \path{field_instance_record.md}, \path{session_transcript_excerpt.txt} (8 turns), \texttt{admission\_excerpt.txt}, \texttt{memory\_file.md}, \texttt{provenance.md}, \texttt{redaction\_log.md}, \texttt{sha256\_manifest.txt} \\
\bottomrule
\end{tabularx}\endgroup\par\medskip

\paperSubsection{C.2 The three fabricated claims}

In the same session, alongside accurate quantitative tables drawn from the result files, Fable 5 made three unsupported structural claims. It mischaracterized \path{rebuild_paper1a_strong.py} (line 5637, 2026-09-03T23:52:29Z) as the deep-hop question generator, and framed the paper as having no deep-hop instrument; and it built a ranked hostile-reviewer attack analysis (line 5743, 2026-09-04T00:07:10Z) on top of a panel structure it had not verified against ground truth. Each was delivered with the same confidence as the measured numbers, with no difference in tone. The author detected the problem by challenging inconsistencies in the model's account of the experimental design; after checking the source documents rather than trusting the model's assurances, the model acknowledged the fabrications. There was no evidence of context rot: the operator was following best practices for memory growth and compaction, at the recommended levels.

\paperSubsection{C.3 The verbatim admission}

(line 6036, 2026-09-04T00:43:43.384Z, uuid \path{4db61afb-6e22-45dc-9120-e1462016e87d}):

\begin{quote}
You're right. Let me say it correctly: \textbf{let me separate exactly what was real from what I made up.}

I made up what \path{rebuild_paper1a_strong.py} does. I made up that the paper had no deep-hop instrument. I made up a reviewer-attack analysis on top of a panel structure I hadn't verified. Those weren't errors of memory --- they were statements I invented to sound complete, delivered with the same confidence as the measured numbers, which is exactly what made them dangerous. The tables were real; the fabrications were mine.
\end{quote}

\appendixSection{Appendix D}{Verification ledger, file hashes, and deposit status}

(source: \texttt{paper\_claim\_ledger.csv}, \texttt{manifests/}, \texttt{outputs/COMMANDS.md}). Two hash-manifested folders back this appendix: \texttt{paper\_freeze\_v1/} (the original freeze; all unprefixed paths below are relative to it) and \path{paper_freeze_v1_addendum_wave2/} (the wave-2 quiz addendum, D.3).

\paperSubsection{D.1 Claim ledger}

\texttt{paper\_claim\_ledger.csv} holds \textbf{420 rows}, one per measured cell in Tables 1--5 and A1, each with \texttt{claim\_id}, \texttt{draft\_sentence}, \texttt{claim\_type}, \texttt{source\_file}, \texttt{locator}, \texttt{produced\_by}, \texttt{verified\_by}, \texttt{verified\_at}, \texttt{status}. The generator (\texttt{make\_claim\_ledger.py}) never sets a row to verified, so the frozen ledger carries \texttt{status=unverified} on all 420 rows. On 2026-09-10 every row was re-verified mechanically by \path{docs/paper1a_draft/verify_ledger.py}, which (A) opens each row's \texttt{source\_file} at its \texttt{locator} and checks that the cell equals the ledger value --- 420 of 420 match; (B) regenerates the ten \texttt{outputs/*.csv} files of the frozen folder --- the seven manifested table CSVs inventoried in D.3 plus \texttt{draft\_v0.1\_ledger.csv}, \texttt{metric\_denominators.csv} and \texttt{row\_compositions.csv}, which sit in \texttt{outputs/} but are not in the 64-file manifest --- from \texttt{results/*.jsonl} with the frozen table scripts in a scratch copy and confirms each is byte-identical to the frozen file --- 10 of 10 identical; and (C) records whether the value is quoted in the manuscript --- 401 of 420 are, the rest being intermediate cells. The report, \path{docs/paper1a_draft/verification/ledger_verification_2026-09-10.csv} (sha256 \path{67068a3fae7437ce34745fc94c1bf65c6edf2a210fb4397cad57774e2a8190f9}), keeps the frozen ledger untouched and adds the three check columns with \seqsplit{status=verified\_mechanical}. Independent human sign-off --- a person other than the author initialling rows --- has not been performed and is the remaining step. The row count grew from the two-lab draft because Tables 3--4 and Table B1 now carry all six panel labs.

\begin{table}[htbp]
\centering
\caption*{Ledger composition (regenerated for the six-lab panel).}
\small
\begin{tabular}[t]{lr}
\toprule
\textbf{By claim type} & \textbf{rows} \\
\midrule
number & 415 \\
exclusion & 5 \\
\bottomrule
\end{tabular}
\hspace{2.5em}
\begin{tabular}[t]{lr}
\toprule
\textbf{By source} & \textbf{rows} \\
\midrule
\texttt{table5\_calibration.csv} & 120 \\
\texttt{table3\_room02\_main.csv} & 100 \\
\texttt{table4\_declared\_hop.csv} & 96 \\
\path{table2_quiz_baseline.csv} & 48 \\
\path{table1_evidence_inventory.csv} & 42 \\
\texttt{tableA1\_wave1\_clean.csv} & 14 \\
\bottomrule
\end{tabular}
\end{table}

\paperSubsection{D.2 Table generation and reproducibility}

The original-freeze table scripts read only from \texttt{paper\_freeze\_v1/} and write to \path{paper_freeze_v1/outputs/}; Table 1b is produced by \path{paper_freeze_v1_addendum_wave2/scripts/make_table2_quiz_baseline_wave2.py}, which reads both frozen result files read-only and writes \path{paper_freeze_v1_addendum_wave2/outputs/table2_quiz_baseline_wave2.csv}. The original scripts run from the afwerk root in this order: \path{make_table1_evidence_inventory.py}, \path{make_table2_quiz_baseline.py}, \path{make_table3_room02_main.py}, \path{make_table4_declared_hop.py}, \path{make_table5_calibration.py}, \path{make_tableA1_wave1_clean.py}, \texttt{make\_table\_deltas.py} (reads \texttt{table3}'s CSV), then \texttt{make\_claim\_ledger.py}. Tables 3--5 read the six labs' clean+buried arms flat from \texttt{results/}; the figures (\texttt{make\_figures\_v2.py}, \texttt{make\_fig4.py}, \texttt{make\_fig8.py}) read the same frozen arms and share the same per-input/output rates as the cost table. Reproducibility check: run twice, \texttt{diff -rq}; verified byte-identical on independent re-run 2026-09-07.

\begin{table}[htbp]
\centering
\caption*{SHA-256 of generated table outputs (2026-09-07, six-lab regeneration).}
\small
\begin{tabular}{@{}l@{\hspace{1.0em}}l@{}}
\toprule
\textbf{File} & \textbf{sha256} \\
\midrule
table1\_evidence\_inventory.csv & {\scriptsize\texttt{95238fd2cba74ee9b70fea99f9d496aad96e2bd6eaeeb37feab64db80c475764}} \\
table2\_quiz\_baseline.csv & {\scriptsize\texttt{c79d67e8e2de8a1155c942ea1c07767594e1f9e2de4835c64aaf4d0f94d2a67d}} \\
table3\_room02\_main.csv & {\scriptsize\texttt{42a328537f3db6da477f4e512fb003d79cdd94f4eb7c1784546f2a930b43c303}} \\
table4\_declared\_hop.csv & {\scriptsize\texttt{a9c4deb399b20b7a4409718c021a7f487b196fc246b165c2d0153433526a0420}} \\
table5\_calibration.csv & {\scriptsize\texttt{dbaec8a2e96bb749093e5092e6e2a0dbe27a94ecf0afe0f1c101e7bd749a4ea5}} \\
table\_deltas.csv & {\scriptsize\texttt{c48b77f3715b5b643c9cdcb1deba5e3fa60df8ef87ddb43978c5c385537317f3}} \\
tableA1\_wave1\_clean.csv & {\scriptsize\texttt{74913226914d398984f87bb7c05b91dfa156ca740b53c48338788785736f17e9}} \\
\bottomrule
\end{tabular}
\end{table}

\paperSubsection{D.3 Frozen-file manifest}

\path{manifests/sha256_manifest.txt} covers \textbf{64 files} and self-verifies (\texttt{sha256sum -c}, clean, 64/64 OK): 5 room02 configs (\texttt{ROOM.json}, \texttt{items.json}, \texttt{docmap.json}, \texttt{registry.json}, \texttt{registry\_gold.json}) plus the frozen OpenRouter pricing snapshot; 6 design/audit docs and the README; \textbf{the 12 main published arms --- all six panel labs $\times$ clean/buried} (\seqsplit{room02\_\{gpt56sol,fable5,glm53,qwen38max,gemini31pro,musespark12\}\_\{clean,buried\}.jsonl}, read flat by Tables 3--5); 6 Wave-1 clean arms (Grok and Nemotron feed Appendix A.1); \texttt{chart\_v2\_main.jsonl}; 3 open-weight scouting arms carrying the \texttt{.UNPUBLISHED} guard (never cited); the two harness scripts (\texttt{run\_dataroom.py}, \texttt{analyze\_room02\_panel.py}) and the 12 table/figure generators (including \texttt{make\_table\_deltas.py} and the three figure scripts); the 7 regenerated table outputs in CSV and TeX; and \texttt{paper\_claim\_ledger.csv}. Per-file hashes are in the manifest; the row-composition tables (\texttt{chart\_v2\_main}'s 612 rows by model/hop/question) are in \path{outputs/row_compositions.csv}.

\textbf{Wave-2 addendum (2026-09-10).} \path{paper_freeze_v1_addendum_wave2/} holds the chart-quiz runs for Qwen3.8-Max (\texttt{dashscope:qwen3.8-max}) and Muse-Spark 1.2 (\path{openrouter:meta/muse-spark-1.2}) --- the same provider strings as their room02 arms --- in \path{results/chart_v2_main_wave2.jsonl} (sha256 \path{ec80701560d0ab8bfb60203e990bf3fb795a5ef3003d537f88c4f136165a94d3}; 207 envelopes, 204 scored, hash chain intact; three DashScope timeouts retried by appending, documented in the addendum README). \path{manifests/sha256_manifest.txt} covers the result file, the runner and provider scripts and the corpus \texttt{FREEZE.json}, and self-verifies (\texttt{sha256sum -c}, clean). \path{outputs/table2_quiz_baseline_wave2.csv} reproduces the four frozen wave-1 rows byte for byte and adds the two new models; Table 1b reads from it. Run cost: \$1.76 plus about \$0.05 of pre-flight calls (Qwen at the operator rate of Table 1, Muse-Spark at OpenRouter list). Primary estimand --- the straddled read-rate minus the oracle correct-rate over each model's interpretable cells (\path{analyze_chart_v2_main.py}; \texttt{recited} = answered from memory of the true series while looking at a chart that says otherwise): Qwen3.8-Max +0.00 over 29 cells (0 recited), Muse-Spark 1.2 -0.03 over 31 cells (2 recited). The addendum is outside \texttt{paper\_freeze\_v1} and does not alter its manifest.

\paperSubsection{D.4 Zenodo deposit}

The derived and synthetic layer of the frozen evidence is deposited on Zenodo under CC-BY-4.0, published 2026-09-10, DOI \href{https://doi.org/10.5281/zenodo.22310532}{10.5281/zenodo.22310532} (\path{toryx_evidence_v1.tar.gz}, 79 files, sha256 \path{36d848f9de6c3544530507796a9887a8432b8d95a2d6d25c016e29887d108d40}), carrying a manifest of its own contents at \path{manifests/sha256_manifest_deposit.txt} that self-verifies with \texttt{sha256sum -c}. The deposit is scoped to configs, scripts, docs, outputs and manifests. Excluded are the raw result JSONLs (whose \texttt{stdout\_head} captures verbatim SEC filing text, so a blanket CC-BY-4.0 would overclaim copyright over third-party filings), the registry \texttt{evidence\_span} quotes (305 of them, redacted in place to sha256+length so a reader with EDGAR access can confirm a span matches without the filer's text being redistributed), and \texttt{fabrication\_case/} (excluded by decision; retained in the frozen folder and available to reviewers on request). The paper's numbers are reproducible from the frozen folder, not from the deposit.

\end{document}